\documentclass[journal]{IEEEtran}
\IEEEoverridecommandlockouts
 \usepackage[english]{babel}

\newcommand{\avec}{{\bf{a}}}

\newcommand{\yvec}{{\bf{y}}}
\newcommand{\uvec}{{\bf{u}}}

\newcommand{\xvec}{{\bf{x}}}

\newcommand{\vvec}{{\bf{v}}}

\newcommand{\onevec}{{\bf{1}}}
\newcommand{\zerovec}{{\bf{0}}}

\newcommand{\Amat}{{\bf{A}}}
\newcommand{\Bmat}{{\bf{B}}}

\newcommand{\Imat}{{\bf{I}}}
\newcommand{\Lmat}{{\bf{L}}}

\newcommand{\Smat}{{\bf{S}}}

\newcommand{\Rmat}{{\bf{R}}}

\newcommand{\Vmat}{{\bf{V}}}
\newcommand{\Wmat}{{\bf{W}}}

\newcommand{\define}{\stackrel{\triangle}{=}}

\newcommand{\be}{\begin{equation}}
\newcommand{\ee}{\end{equation}}
\newcommand{\beqna}{\begin{eqnarray}}
\newcommand{\eeqna}{\end{eqnarray}}

\usepackage{cite}
\usepackage{amsmath,amssymb,amsfonts}
\usepackage{xcolor}
\usepackage{subcaption}
\usepackage{algorithm,algorithmic}
\usepackage{acronym}
\usepackage{enumitem}
\usepackage{ntheorem}
\newtheorem{theorem}{Theorem}
 \newtheorem{definition}{Definition}
  \newtheorem{proposition}{Proposition}
    \newtheorem{claim}{Claim}
\definecolor{NewColor}{rgb}{0,0,0}

\acrodef{admm}[ADMM]{alternating
direction method of multiplier}
\acrodef{ac}[AC]{alternating current}
\acrodef{arma}[ARMA]{auto-regressive moving average}
\acrodef{cdf}[CDF]{cumulative distribution function
}
\acrodef{cfar}[CFAR]{constant false alarm rate}
\acrodef{crb}[CRB]{ Cram$\acute{\text{e}}$r-Rao  bound}
\acrodef{dau}[DAU]{deep algorithm unrolling}
\acrodef{dc}[DC]{direct current}
\acrodef{dnn}[DNN]{deep neural network}
\acrodef{dsp}[DSP]{digital signal processing}
\acrodef{ems}[EMS]{energy management system}
\acrodef{evd}[EVD]{eigenvalue decomposition}
\acrodef{gan}[GAN]{generative
adversarial network}
\acrodef{gsp}[GSP]{graph signal processing}
\acrodef{gnn}[GNN]{graph neural network}
\acrodefplural{gnn}[GNN-based]{graph neural network-based}
\acrodef{glrt}[GLRT]{generalized LRT}
\acrodef{gso}[GSO]{graph shift operator}
\acrodef{gft}[GFT]{Graph Fourier Transform}
\acrodef{gmrf}[GMRF]{Gaussian Markov random field}
\acrodef{gcs}[GCS]{Graph Convolutional Skip}
\acrodef{gcn}[GCNN]{Graph Convolutional Neural
Network}
\acrodef{hpf}[HPF]{high-pass filter}
\acrodef{iid}[i.i.d.]{independent and identically distributed}
\acrodef{ista}[ISTA]{iterative soft thresholding algorithm}
\acrodef{kkt}[KKT]{Karush-Kuhn-Tucker}
\acrodef{lrt}[LRT]{likelihood ratio test}
\acrodefplural{LRTs}{likelihood ratio tests}
\acrodef{lista}[LISTA]{learned \ac{ista}}
\acrodef{lmmse}[LMMSE]{linear minimum mean-squared error}
\acrodef{lpf}[LPF]{low-pass graph filter}
\acrodef{lti}[LTI]{linear-time invariant}
\acrodef{map}[MAP]{maximum a-posteriori probability}
\acrodef{ml}[ML]{maximum likelihood}
\acrodef{mse}[MSE]{mean-squared error}
\acrodef{mmse}[MMSE]{minimum MSE}
\acrodef{pdf}[PDF]{probability density function}
\acrodef{pf}[PF]{power flow}
\acrodef{psd}[PSD]{positive semi-definite}
\acrodef{psse}[PSSE]{power systems state estimation}
\acrodef{roc}[ROC]{Receiver Operating Characteristic}
\acrodef{rbf}[RBF]{radial basis function}
\acrodef{snr}[SNR]{signal-to-noise ratio}
\acrodef{svd}[SVD]{singular value decomposition}
\acrodef{scada}[SCADA]{supervisory control and data acquisition}
\acrodef{tv}[TV]{total variation}
\acrodef{wls}[WLS]{weighted least squares}
\acrodef{wrt}[w.r.t.]{with respect to}
\acrodef{wsn}[WSN]{wireless sensor network}

\usepackage[most]{tcolorbox}
\newtcolorbox{boxA}[2][]{%
  attach boxed title to top center
              = {yshift=-8pt},
  colback      = blue!5!white,
  colframe     = blue!75!black,
  fonttitle    = \bfseries,
  colbacktitle = blue!85!black,
  title        = #2,#1,
  enhanced,
}
\begin{document}

\title{Verifying the Smoothness of Graph Signals: A Graph Signal Processing Approach }
\author{\IEEEauthorblockN{Lital Dabush \IEEEmembership{Student~Member,~IEEE},  and Tirza Routtenberg \IEEEmembership{Senior~Member,~IEEE}}
\thanks{Lital Dabush and Tirza Routtenberg are 
with the School of Electrical and Computer Engineering, Ben-Gurion University of the Negev
 Beer-Sheva 84105, Israel.  e-mail: litaldab@post.bgu.ac.il,~tirzar@bgu.ac.il. 
 This research was supported by the Israel Science Foundation (Grant No. 1148/22) and by the Israel Ministry of National Infrastructure and Energy.  L. Dabush is a fellow of the AdR Women Doctoral Program.
 }
}
\maketitle
\begin{abstract}
Graph signal processing (GSP) deals with the representation, analysis, and processing of structured data, i.e. 
{\em{graph signals}} that are defined on the vertex set of a generic graph.
A crucial prerequisite for applying various GSP  and graph neural network (GNN) approaches is that the examined signals are {\em{smooth graph signals}}  with respect to the underlying graph, or, equivalently, have low graph total variation (TV). 
In this paper, we develop GSP-based approaches to verify the validity of the smoothness assumption of given signals (data) and an associated graph. The proposed approaches are based on the representation of network data as the output of a graph filter with a given graph topology.
In particular, we develop two smoothness detectors for the graph-filter-output model: 1) the likelihood ratio test (LRT) for known model parameters; and 2) a semi-parametric detector that estimates the graph filter and then validates its smoothness.
The properties of the proposed GSP-based detectors are investigated, and some special cases are discussed.
The performance of the GSP-based detectors is evaluated {using synthetic data, data from the IEEE 14-bus power system, and measurements from a network of light intensity sensors}, 
 under different setups. The results demonstrate the effectiveness of the proposed
approach and its robustness to different generating models, noise levels, and number of samples.
\end{abstract}
\section{Introduction}
A growing trend in signal processing and machine learning is to develop theories and models for analyzing data defined in
irregular domains, such as graphs. 
In particular,  the emerging field of  \ac{gsp} extends classical signal processing methodologies, such as filtering and sampling, to the graph domain \cite{Sandryhaila_Moura_2014,Shuman_Ortega_2013,8347162,Milos,kroizer2022routtenberg,miettinen2021graph}. 
Graphs often express relational ties, such as social, economic, and biological networks, for
which several mathematical and statistical models relying on graphs have been proposed  \cite{Kolaczyk,Huang2018}.
Another case is that of physical infrastructure networks, such as utility and transportation
networks \cite{7472903}, 
where physical laws and connectivity patterns define the graph structure behind the signals.

A crucial prerequisite for applying various \ac{gsp}  and \ac{gnn} approaches is that the examined signals are smooth graph signals  \ac{wrt} the underlying graph or, equivalently, have low graph \ac{tv}.
Specifically, a graph signal is considered smooth when the signal values associated with the two end nodes (vertices) of edges with large weights tend to be similar. 
This smoothness property of graph signals has been used for various {{tasks}}, such as
denoising with Laplacian-smoothness regularization \cite{Laplacianregularization, Dabush2021Routtenberg, rudin1992nonlinear,zhu2006graph}, data classification \cite{Sandryhaila_Moura_2014},  {{sampling and 
 recovery \cite{7352352,6854325,smoothVSbandlimited, reconstruction_of_smooth_graph_signals}, and anomaly detection \cite{ortega2022introduction,verdoja2020graph}. These tools have been applied in various applications, including sensor networks \cite{ortega2022introduction}, finance \cite{9443573}, social networks \cite{bandlimite_PSSE}, traffic flow prediction \cite{bandlimite_PSSE}, and power systems \cite{Dabush2021Routtenberg}. 
For example, 
in power systems, the smoothness of system states (voltages)  is utilized for the detection of hidden false data injection attacks \cite{bandlimite_PSSE,drayer2018detection,dabush2022Routtenberg_detection,morgenstern2023protection}. 
In social networks, the prediction of individual properties (e.g. political opinions) relies on the assumption that connected entities (``friends") share similar characteristics, demonstrating smoothness concerning the social graph \cite{bandlimite_PSSE}, \cite{Dong_Vandergheynst_2016}. 
In this context, signal prediction and denoising based on high-pass graph filter regularization approaches \cite{Rama2020Anna} is not valid if the smoothness assumption does not hold. 
Another example is the detection of a malfunctioning sensor in sensor networks \cite{Sandryhaila_Moura_2014}, where the assumption that close sensors measure similar temperatures (i.e. smoothness concerning the graph's topology) is essential. 
}}Similarly, in \acp{gnn} node classification \cite{kipf2016semi}, it is assumed that neighboring nodes tend to have the same labels, i.e. they are smooth graph signals. 
{However, if the processed signals are not smooth \ac{wrt} the given topology, the performance of the associated task may degrade severely. }
Despite the prevalent assumption of signal smoothness in both \ac{gsp} and \ac{gnn} throughout numerous studies, there is a lack of rigorous validation of this property, where it is often 
treated as an axiom or an implicit assumption.
 This validation is particularly relevant in physical applications, where real-world data may exhibit diverse characteristics, and underscores the need for careful validation of signal smoothness assumptions.

In order to study graph signals and implement \ac{gsp} tools, signals measured over the nodes of a network can be modeled as the outputs of graph filters \cite{Rama2020Anna,Yiran2022Hoi_To, Schultz2021Marisel,kroizer2022routtenberg,isufi2022graph}.
Similar to linear-time-invariant  filters, graph filters can be classified as low-pass, band-pass, or high-pass, according to their frequency responses computed through the \ac{gft} \cite{Sandryhaila_Moura_2014}. 
Graph low-pass filters (LPFs) and signals are strongly connected to smooth graph filters \cite{zhu2006graph, WANG2016119}. 
However, utilizing graph LPFs \cite{7208894, bandlimite_PSSE} requires choosing the filter's order and typically involves an \ac{evd}  of the Laplacian,
which is computationally intensive. 
Thus, in practice, relying upon an assumed smoothness property is preferable for various tasks (see, e.g. \cite{zhu2006graph,rudin1992nonlinear, Perraudin2017Vandergheynst,7352352,6854325,verdoja2020graph}). 

In the case of an unknown topology, there are many topology-learning works that are based on the assumption of smooth data  \ac{wrt} the topology \cite{Dong_Vandergheynst_2016, Vassilis_2016,ortega2017, stankovic2020graph,shafipour2021identifying, pasdeloup2017characterization}. However,   these works do not validate the smoothness assumption. In \cite{He2012Hoi-To}, the authors suggest a first-order low-pass graph signal detector for datasets whose social/physical models are unknown. However, the detector cannot be modified to higher-order low-pass graph signals.
Furthermore, as it lacks knowledge of the graph, it requires a significantly larger amount of data to accurately estimate the covariance matrix of the measurements, which may be impractical.
In \cite{hu2013matched}, matched signal detection on graphs is derived based on the smoothness prior in order to decide which graph structure a signal is more likely to be embedded in. In \cite{Isufi2018Leus}, new methods to detect blind topology changes in a graph are developed based on testing the smoothness/sparsity level of the measured signal \ac{wrt} the original graph. 
In the case of a known graph topology, there are several works
that test the validity of the smoothness assumption for specific applications. For example, 
in
\cite{drayer2018detection,Dabush2021Routtenberg,2021Anna},  the smoothness assumption of the voltages in power systems is investigated. 
Similarly, in \cite{Schultz2021Marisel}, the smoothness of data from biological organisms is validated empirically 
for the application of the detection of malfunctioning or nefarious agents. {In \cite{preti2019decoupling}, the brain  function's smoothness \ac{wrt} the anatomical structure is explored.}
However, these validations are specific to the application tested.
Thus, a validation of the smoothness assumption for general applications with unknown graph signal generation processes has not yet been demonstrated nor analyzed.

In this paper, we address the question of whether a given set of graph signals is smooth \ac{wrt} a given graph, through new \ac{gsp}  methods. 
First, we represent the data obtained
from a network process as the output of a smooth graph filter.
To this end, we define and exemplify the concept of smooth graph filters.
We show that under the model of graph-filtered white Gaussian noise, covariance-based detectors capture signal smoothness. 
  Then, we develop two covariance-based \ac{gsp} detectors: 1)  the \ac{lrt} for smooth graph signals, where all model parameters are assumed to be known;
and 2) a semi-parametric detector that estimates the graph filter that generates the data and then tests if it is a smooth graph filter. This detector does not impose any assumptions on the underlying data-generating model, making it suitable for a wide range of graph signals.
The proposed detectors are implemented within the node domain,  which obviates the necessity of performing \ac{evd} and enables distributed computation.
A method for setting the threshold by the required probability of false alarm of the detectors is provided.
The performance of the proposed detectors is evaluated using synthetic data,  power system data, {and measurements from light intensity sensors}, in terms of probability of detection, sensitivity to the number of samples, the smoothness level of the graph filters, and robustness to scaling.
The simulations demonstrate that the \ac{gsp}-based methods are capable of detecting smooth graph signals and outperform existing detectors.
 
The remainder of the paper is organized as follows.
 In Section \ref{GSP_sec}, we present a background on \ac{gsp} and the observation model.
 Section \ref{sub_smooth_detect} describes the modeling of the problem and the two \ac{gsp} detectors.
In Section \ref{sec;sim}, a simulation study is presented. Finally, the conclusions appear in Section \ref{Conclusions}.

In the rest of this paper, vectors and matrices are denoted by boldface lowercase letters and boldface uppercase letters, respectively. 
The notations $(\cdot)^T$, $(\cdot)^{-1}$, $(\cdot)^{\dagger}$, $|\cdot|_+$
and $\text{Tr}(\cdot)$ denote the transpose, inverse, Moore-Penrose pseudo-inverse, pseudo-determinant (i.e. the product of the nonzero eigenvalues of a 
matrix), and trace operators,
respectively.
The $m$th element of the vector $\avec$  and the $(m, q)$th element of the matrix $\Amat$ are denoted by $a_m$ and $A_{m,q}$, respectively. 
  $\Imat$, and
 $\onevec$ and $\zerovec$ denote the identity matrix, and
 vectors of ones and zeros, respectively, with appropriate dimensions, and  
 $||\cdot||$ denotes the Euclidean $l_2$-norm of vectors. 
For a vector $\avec$, ${\text{diag}}(\avec)$ is a diagonal matrix whose $(n,n)$th entry is $a_n$; when applied to a matrix, ${\text{diag}}(\Amat)$ is a vector with the diagonal elements of $\Amat$.

\section{Background and model}
\label{GSP_sec}
In this section, we  present the background for \ac{gsp} in Subsection \ref{Graph Signal Processing Overview} and the smooth graph filter definition in Subsection \ref{Smooth graph filter}. Finally, we describe the considered measurement model as an output of a smooth \ac{gsp} filter in Subsection \ref{subsec_measurements_model}.

\subsection{Background: Graph Signal Processing (\ac{gsp})} \label{Graph Signal Processing Overview}
Let ${\mathcal{G}}({\mathcal{V}},\xi)$ be a general undirected  weighted graph,
where ${\mathcal{V}}=\{1,\ldots,N\}$ and ${\xi}$ are the sets of nodes and edges, respectively.
The matrix $\Wmat\in{\mathbb{R}}^{N\times N}$ is the weighted adjacency matrix of the graph $\mathcal{G}({\mathcal{V}},\xi)$, where $W_{k,n}\geq 0$ denotes the weight of the edge between node $k$ and node $n$, and 
$W_{k,n}= 0$ if no  edge exists between $k$ and $n$.
The graph Laplacian matrix is a real positive semi-definite matrix with the  \ac{evd} defined as 
\be
\Lmat\define {\text{diag}}(\Wmat\onevec) - \Wmat=\Vmat{\text{diag}}(\pmb{\lambda}) \Vmat^{-1},   \label{SVD_new_eq}
\ee
where the columns of $\Vmat$ are the eigenvectors of $\Lmat$, $\Vmat^{T}=\Vmat^{-1}$, and $\pmb{\lambda} \in {\mathbb{R}}^{N}$ is a vector of the ordered eigenvalues of 
$\Lmat$.
We assume that ${\mathcal{G}}({\mathcal{V}},\xi)$ is a connected graph, and thus, $\lambda_2\neq 0$. 
By analogy to the  frequency of signals in \ac{dsp},
 the Laplacian eigenvalues,
$\lambda_1,\ldots,\lambda_N$, can be interpreted as the graph frequencies.
Together with the eigenvectors in
$\Vmat$, they define the spectrum of the graph 
\cite{Shuman_Ortega_2013}.

A graph signal is a function that resides on a graph, assigning a scalar value to each node.
The \ac{gft} of a graph signal $\avec\in{\mathbb{R}}^N$  \ac{wrt} the graph ${\mathcal{G}}({\mathcal{V}},\xi)$ is defined as  \cite{Shuman_Ortega_2013,Sandryhaila_Moura_2014}
 \begin{equation}
\label{GFT}
\tilde{\avec}\triangleq \Vmat^{-1}\avec.  
 \end{equation}
  Similarly,  the inverse \ac{gft} is obtained by a left multiplication of $\tilde{\avec}$ by $\Vmat$.
The \ac{tv} of a graph signal, $\avec$, is defined as 
\begin{equation}
\label{eq:Dirichlet energy}
{\avec^T}\Lmat\avec 
=%
\frac{1}{2} \mathop \sum _{k=1}^N\sum_{n=1}^N W_{k,n}\big(a_k - a_n\big)^2
=\sum_{n=1}^N\lambda_n\tilde{a}_n^2,
\end{equation} where  the first equality is obtained by substituting  \eqref{SVD_new_eq} and the second equality is obtained 
by substituting \eqref{SVD_new_eq} and \eqref{GFT}.
 
The \ac{tv} from \eqref{eq:Dirichlet energy}  
is a smoothness measure, which is used in graphs to quantify changes \ac{wrt} the variability that is encoded by the weights of the graph \cite{Shuman_Ortega_2013}.
{The \ac{tv} can also be defined based on the $\ell_1$-norm \cite{Sandryhaila_Moura_2014}, which is less commonly used than the $\ell_2$-norm due to traceability issues.} 
A graph signal, $\avec$, is smooth if $\avec^T\Lmat\avec\leq\varepsilon$, 
where $\varepsilon$ is small in terms of the specific application  \cite{Shuman_Ortega_2013}.
Thus, the smoothness assumption requires connected nodes to have similar values, 
and the graph signal spectrum to be in the small eigenvalues region (see \eqref{eq:Dirichlet energy}).

Linear and shift-invariant graph filters play essential roles in \ac{gsp}.
These filters generalize linear time-invariant filters 
used 
in \ac{dsp}, and enable processing over graphs \cite{8347162,Shuman_Ortega_2013}.
A Laplacian-based graph filter can be defined in the graph frequency domain as  a function $h(\cdot)$ that  allows an \ac{evd}  \cite{8347162}:
\begin{equation}\label{graph_filter} h(\Lmat)\hspace{-0.05cm}\define\hspace{-0.05cm}\Vmat{\text{diag}}(h(\pmb{\lambda}))\Vmat^{-1},~h(\pmb{\lambda})=[h(\lambda_1),\dots,h^T(\lambda_{N})], \hspace{-0.05cm}\end{equation}
  where   $h(\lambda_n)$ is the graph  filter  frequency response  at the graph frequency $\lambda_n$, $n=1,\ldots,N$.
  It should be noted that if an eigenvalue of $\Lmat$ has a multiplicity
greater than 1, then its corresponding graph frequency responses should be identical for all the equal eigenvalues (see \cite{ortega2022introduction} Chapter 3). That is, 
\be 
\label{constraint1}
h(\lambda_m) = h(\lambda_k),~~~ \forall \lambda_m=\lambda_k.
\ee

A graph filter applied on a graph signal is a linear operator   that satisfies the following:
\begin{equation}
\label{a_out_a_in}
   \avec^{(\rm{out})}=h(\Lmat)\avec^{(\rm{in})},
\end{equation}
where $ \avec^{(\rm{out})}$ and $\avec^{(\rm{in})}$ are the output and input graph signals.

\subsection{Smooth Graph Filters}\label{Smooth graph filter}
The objective of this research is to develop a method to validate the smoothness of signals measured over networks. 
To this end,
we model smooth graph signals as the output of a smooth graph filter.
In this subsection, we first establish a clear definition of a smooth graph filter and provide relevant examples.
The following definition of smooth graph filters, which is a normalized version of   the definition in 
\cite{shaked2021identification},  implies that for smooth graph filters, the output graph signal is smoother than the input graph signal, on average.
\begin{definition}\label{smoothdef}
Let the elements of the input graph signal, $\avec^{(\rm{in})}$, be \ac{iid}  
 zero-mean random variables. Then $h(\cdot)$ in  \eqref{graph_filter} is
a smooth graph filter if 
  \beqna\label{SGF_normalized}
     r\define\frac{{\rm{E}}[\|\avec^{(\rm{in})}\|^2]}{{\rm{E}}[\|\avec^{(\rm{out})}\|^2]} \times
     \frac{{\rm{E}}[(\avec^{(\rm{out})})^T\Lmat\avec^{(\rm{out})}]}{{\rm{E}}[(\avec^{(\rm{in})})^T{\Lmat}(\avec^{(\rm{in})})]}< 1,
 \eeqna
 where 
$\avec^{(\rm{out})}$ is given in \eqref{a_out_a_in}.
\end{definition}
 This definition is based on the graph \ac{tv}, $\avec^T\Lmat\avec$, defined in (\ref{eq:Dirichlet energy}), { and can also be defined using the $\ell_1$-based \ac{tv}.}
 By using \eqref{GFT}, \eqref{eq:Dirichlet energy}, and the fact that the Euclidean norm  is invariant under multiplication by  unitary matrices (here, by $\Vmat^{-1}$),  we obtain that $r$ from \eqref{SGF_normalized} can be written as
\beqna\label{eq_calc_r_mid_step}
 r=\frac{{\rm{E}}[\|\tilde{\avec}^{(\rm{in})}\|^2]}{{\rm{E}}[\|\tilde{\avec}^{(\rm{out})}\|^2]} \times
     \frac{{\rm{E}}[\sum\nolimits_{n=1}^N\lambda_n (\tilde{a}^{(\rm{out})}_n)^2]}{{\rm{E}}[\sum\nolimits_{n=1}^N\lambda_n (\tilde{a}^{(\rm{in})}_n)^2]}.
\eeqna
By substituting 
\eqref{GFT}, \eqref{graph_filter}, and \eqref{a_out_a_in}
in \eqref{eq_calc_r_mid_step}, and
under the assumption that the elements of $\avec^{(\rm{in})}$ are \ac{iid} variables, the condition in \eqref{SGF_normalized} can be written in the graph frequency domain as
 \beqna\label{eq_SGF_gft}
      r=\lambda_{\rm{avg}}^{-1} \times \frac{\sum_{n=1}^N{\lambda_n h^2({\lambda}_n)}}{\sum_{n=1}^N{ h^2({\lambda}_n)}}< 1,~~\lambda_{\rm{avg}}\define \frac{1}{N}\sum_{n=1}^N\lambda_n.
 \eeqna  
  That is, $r$ is the ratio of the weighted average of $\pmb{\lambda}$, weighted by $h^2(\pmb{\lambda})$, and the algebraic average. Thus, if the energy of the frequency response is uniformly distributed across all graph frequencies, the ratio will be  $1$, indicating that the graph filter is not smooth. If the energy is biased towards low (high) graph frequencies, the ratio will be lower (greater) than $1$.

Definition \ref{smoothdef} has the desired property of being scale-invariant, meaning that multiplying the graph filter, $h(\pmb{\lambda})$, by a constant $\alpha\neq 0$ will not change the value of $r$ and the graph filter smoothness as defined in  \eqref{SGF_normalized}.
 Moreover, the definition of $r$ in \eqref{eq_SGF_gft} implies that the graph filter smoothness is solely determined by the amplitude of $h(\cdot)$, and not by the sign of the elements of $h(\cdot)$, similar to in conventional \ac{dsp} theory, where the description of signals as high- or low-pass is solely determined by the amplitude of the Fourier transform.

In Box 1, we present the normalized versions of the graph frequency response, $ h(\lambda)$,  of smooth graph filters that are commonly used in the \ac{gsp} and networks literature.
The normalization factor of the graph filters, $\beta$, can be arbitrarily set, where in the literature, usually $\beta=1$. In this paper, we use $\beta_{GMRF}$, $\beta_{Tikh}$, and $\beta_{Diff}$ such that
$\sum_{n=1}^Nh(\lambda_n)^2=N$.
\begin{boxA}[colback=blue!5!white]{Box 1: Examples of smooth graph filters}
\begin{enumerate}[wide, labelwidth=!, labelindent=0pt]
    \item \ac{gmrf} with a Laplacian precision matrix \cite{Dong_Vandergheynst_2016,Vassilis_2016}:
    \begin{align}\label{GMRF filter}
 h_{\text{GMRF}}(\lambda)=\beta_{\text{GMRF}}\begin{cases}\frac{1}{\sqrt{\lambda}} \quad &\lambda\neq0\\
    0 \quad& \lambda=0\end{cases},\end{align}
    
    \item  Laplacian (Tikhonov) Regularization \cite{Vassilis_2016,isufi2022graph}:
    \begin{align}\label{Tikhonov filter}h_{\text{Tikh}}(\lambda)=\frac{\beta_{\text{Tikh}}}{1+\alpha\lambda},~\alpha>0, \end{align}

   \item Heat Diffusion Kernel \cite{thanou2016learning,Vassilis_2016}:
    \begin{align}\label{diffusion filter}h_\text{Diff}(\lambda)=\beta_\text{Diff}\exp{(-\tau\lambda}),~\tau>0.\end{align}

\end{enumerate} 
\end{boxA}

\begin{figure}[hbt]
    \centering
    \includegraphics[width=0.7\columnwidth]{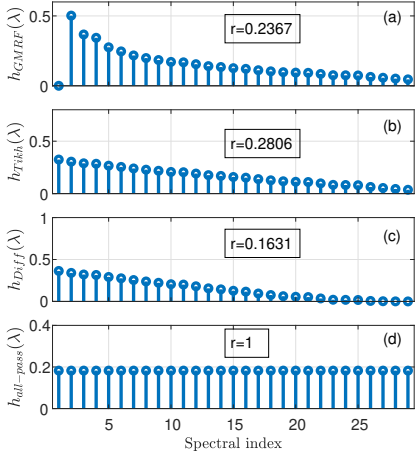}
    \caption{
        The normalized frequency response of: a) the \ac{gmrf} graph filter from \eqref{GMRF filter}; 
    b) the Tikhonov Regularization graph filter from \eqref{Tikhonov filter}, with $\alpha=0.2$; 
    c) the Heat Diffusion Kernel graph filter from \eqref{diffusion filter}, with $\tau=0.1$; 
    and d) the all-pass graph filter, 
    for the graph that represents the IEEE $30$ bus system.}
    \label{fig_filters}
\end{figure}

\begin{theorem}\label{non_increasing_theorem}
{Any monotonically non-increasing graph filter with at least two distinct coefficients is a smooth graph filter according to Definition \ref{smoothdef}.}
\end{theorem}
\begin{IEEEproof}
    {See Appendix \ref{appendix_smoothness_of_box_filters}.}
\end{IEEEproof}
{According to Theorem \ref{non_increasing_theorem},} \eqref{Tikhonov filter} and \eqref{diffusion filter} in Box 1 are smooth graph filters. 
The GMRF graph filter in \eqref{GMRF filter} satisfies $h(\lambda_1)=0$. Thus, while it generally has smooth characteristics, certain extreme topologies can lead to a non-smooth graph filter. Therefore, it is crucial to evaluate Definition \ref{smoothdef} for this filter, considering the specific topology.
In addition,  to demonstrate the effect of the considered  smooth graph filters on the frequency content of an  input graph signal,  Fig. \ref{fig_filters} shows the output graph signals in the graph frequency domain,  i.e. 
   $\Vmat^{-1}\avec^{(\rm{out})}$, that are obtained by applying \eqref{GMRF filter}, \eqref{Tikhonov filter}, \eqref{diffusion filter}, and the non-smooth all-pass graph filter, $h_{\text{all-pass}}(\Lmat)=\Imat$,  on the input graph signal $\tilde{\avec}_{in}=\onevec$ for the graph associated with the IEEE $30$ bus system \cite{iEEEdata}.
It can be seen that for the graph filters from Box 1, the high-frequency components are attenuated compared to the low-frequencies. 
In contrast, there is no change in the frequency contents after applying the all-pass filter. This can also be seen from the $r$ values of the filters { in Fig. \ref{fig_filters} that are obtained by substituting \eqref{GMRF filter}-\eqref{diffusion filter} and the eigenvalues of the associated Laplacian matrix in \eqref{eq_SGF_gft}}.

It should be noted that one could also propose measuring the smoothness of practical networked data using the following definition of a graph \ac{lpf} from \cite{Rama2020Anna}.

\begin{definition}\label{def_LPF}
The graph filter 
 in \eqref{a_out_a_in}
is a  graph \ac{lpf} of order $K$
with a cutoff frequency  $\lambda_K$ 
if  $\eta_K<1$, where 
\begin{equation}
\label{eta_k_def}
    {\eta}_{k}\define \frac{\max\{\vert{h}({\lambda}_{k+1})\vert,\ldots,\vert{h}({\lambda}_{N})\vert\}}{\min\{\vert{h}({\lambda}_{1})\vert,\ldots,\vert{h}({\lambda}_{k})\vert\}},~ k=1,\ldots,N-1.
\end{equation}
\end{definition}
The following claim describes the relationship between Definition \ref{smoothdef} (proposed here) and 
Definition  \ref{def_LPF} (proposed in \cite{Rama2020Anna}).
\begin{claim}\label{claim_equiv_to_lpf}
Let $J\in\{1,\ldots,N-1\}$ be the index for which $\lambda_J\leq\lambda_{\rm{avg}}<\lambda_{J+1}$, where $\lambda_{\rm{avg}}$ is defined after \eqref{eq_SGF_gft}.
Then,
any graph \ac{lpf}  of order $K\leq J$
(according to Definition  \ref{def_LPF}) is a smooth graph filter (according to Definition \ref{smoothdef}) under the condition 
\begin{equation}\label{eq_cond_eta_1}
    \eta_K^2<\frac{\sum_{n=1}^K(\lambda_n-\lambda_{\rm{avg}})}{\sum_{n=1}^J(\lambda_n-\lambda_{\rm{avg}})},
\end{equation}
where $    {\eta}_{K}$ is defined in \eqref{eta_k_def}.
\end{claim}
\begin{IEEEproof}
The proof appears in Appendix \ref{proof_claim_equiv_to_lpf}.
\end{IEEEproof}
For the special case of $K=J$ we obtain that 
 any 
 graph \ac{lpf}  of order $J$ is a smooth graph filter.
Thus, the two definitions coincide if the order of the graph filter is exactly the index below the averaged  Laplacian  eigenvalue. 
The condition in \eqref{eq_cond_eta_1}  ensures that most of the filter energy  is concentrated in the first $J$ graph frequencies. 
According to Claim \ref{claim_equiv_to_lpf}, using Definition \ref{smoothdef} to identify smooth graph filters encompasses all \ac{lpf} graph filters of order $K<J$ s.t. $\eta_K$ satisfies \eqref{eq_cond_eta_1}. Since Definition  \ref{smoothdef} does not {necessitate specifying} the filter order $K$, {nor computing the \ac{gft}, is more robust to outliers in the high graph frequencies region, and achieves better results in the tested simulations (see Section \ref{sec;sim}); hence, it is favored here. }
\subsection{Measurement Model}
\label{subsec_measurements_model}
We 
 consider a graph signal model as an output of a graph filter $h(\Lmat)$, as defined in \eqref{a_out_a_in}, where the input graph signal is white Gaussian noise. 
This model is widely-used for different smooth graph filters to represent network various  signals   (see, e.g. \cite{Vassilis_2016,egilmez2018graph,Dong_Bronstein_Frossard2020,2021Anna}).
Mathematically, the assumed measurement model is given by
\begin{align}\label{model_mu}
    \bar{\xvec}[m]= \pmb{\mu}+\bar{h}(\Lmat)\yvec[m], ~ m = 1 \ldots M,
\end{align}
 where $m$ is a time index and $\pmb{\mu}\in \mathbb{R}^N$ represents a general shift.
The  sequence $\{\yvec[m]\}_{m=1}^M$  is a sequence of  \ac{iid}  Gaussian random vectors,  $\yvec[m] \sim \mathcal{N}(\zerovec,\sigma^2\Imat)$, where $\sigma^2>0$.  
Thus,  the output graph signal, $\{\bar{\xvec}[m]\}_{m=1}^M$, is a sequence of \ac{iid} Gaussian random vectors, $\bar{\xvec}[m] \sim \mathcal{N}(\pmb{\mu},\sigma^2\bar{h}^2(\Lmat))$.

 It should be noted that the values of $\sigma^2$ and $h(\Lmat)$ cannot be jointly (uniquely) determined without additional information.
 This is due to the inherent ambiguity present in the measurements of $\{\bar{\xvec}[m]\}_{m=1}^M$, i.e. the inability to separate $\sigma^2$ and $\bar{h}(\Lmat)$ in the \ac{pdf} of $\{\bar{\xvec}[m]\}_{m=1}^M$.
 In order to simplify the model, we are using the following claim.
\begin{claim}\label{prop_mu}
Assume the following measurement model: 
\begin{align}\label{model}
    \xvec[m]= h(\Lmat)\yvec[m], ~ m = 1 \ldots M,
\end{align}
where $h(\Lmat)$  satisfies
\begin{align}\label{filter_mu}
    h^2(\Lmat)=\bar{h}^2(\Lmat)+\frac{1}{\sigma^2}\pmb{\mu}\pmb{\mu}^T.
\end{align}
Then, 
\begin{enumerate}
\item The model in \eqref{model} is equivalent to the model in \eqref{model_mu} in the sense 
that the expected \ac{tv} of the output graph signals is identical. 
That is, 
\beqna
\label{exp1}
{\rm{E}}[\bar{\xvec}^T[m]\Lmat \bar{\xvec}[m]]={\rm{E}}[ {\xvec}^T[m]\Lmat {\xvec}[m]]
,
\eeqna
$m=1,\ldots,M$, and the  \ac{tv} is not a function of $m$.
\item For $h(\Lmat)$ that satisfies \eqref{filter_mu} to be a valid graph filter, i.e. diagonalizable by $\Vmat$, $h(\Lmat)$ must be symmetric and $\pmb{\mu}$ must be proportional to one of the eigenvectors of $\Lmat$.
 \end{enumerate}
\end{claim}
\begin{IEEEproof}
    The proof appears in Appendix \ref{proof_prop_mu}.
\end{IEEEproof}

Henceforth, for the sake of simplicity and based on Claim \ref{prop_mu}, the measurement model in \eqref{model} is assumed (where ${h}(\Lmat)$ is a graph filter) in order to test the smoothness of the data.
That is,  we implicitly assume that $\pmb{\mu}$ is proportional to an eigenvector of $\Lmat$, $\vvec_n$.
This model implies that $\xvec[m]\sim\mathcal{N}(\zerovec,\sigma^2{h}^2(\Lmat))$.

It can be seen in \eqref{model_mu} and \eqref{model} that in the considered model, the network dynamic determines the covariance matrix of the output graph
 signal. 
In general, the expected smoothness of a graph signal is composed of the smoothness of its mean and the smoothness of its covariance (see more in Appendix \ref{proof_prop_mu}). 
Using the model from \eqref{model}  enables testing the smoothness of the signal by using covariance-based detectors. 
This formulation reduces the number of parameters to be estimated 
and can increase the stability of the analysis compared to the model in \eqref{model_mu}. For example, the \ac{gmrf} graph filter in \eqref{GMRF filter} is a singular matrix, but in the modified model, we have the filter  $h(\Lmat)=\Lmat^{\dagger}+\alpha\onevec\onevec^T$, which is not singular for  $\alpha\not=0$.

To conclude this section, we present the following  statistical proposition, which will be used in Subsections  \ref{sub;detectors_known_model} and \ref{sec;semi_parametric} to analyze the detectors obtained under the  model from \eqref{model}.
\begin{proposition}\label{prop}
Let $h_{a}(\Lmat)$ and  $h_{b}(\Lmat)$ be general graph filters,
and assume that $\xvec[m]\sim \mathcal{N}(\zerovec,h_{b}^2(\Lmat)), ~m=1,\ldots,m$ is a sequence of \ac{iid} Gaussian random vectors.
Then, 
\beqna\label{eq_pfa_general}
  \Pr\left(\sum_{m=1}^M \xvec^T[m] h_{a}(\Lmat)\xvec[m]>\delta\right)\hspace{2cm}\nonumber\\=1-Q\left({M,\{w(\lambda_n)\}_{n \in {\text{supp}}(h_b)}},\delta\right),
 \eeqna
 where  
$  {\text{supp}}(h_b)\define \{n\in\{1,\ldots,N\}| h_{b}(\lambda_n)\neq0\}
$ 
 and $Q$ is the \ac{cdf} of a weighted sum of independent centralized $\chi$-square random variables with $M$ degrees of freedom{, i.e. \beqna Q\left({M,\{w(\lambda_n)\}_{n \in {\text{supp}}(h_b)}},\delta\right) \hspace{3cm}\nonumber\\= \Pr\left(\sum_{n \in {\text{supp}(h_b)}} w(\lambda_n)c_M^{(n)} \leq \delta\right),\eeqna
where \(\{c_M^{(n)}\}_{n \in {\text{supp}(h_b)}}\) are independent $\chi$-square random variables with \(M\) degrees of freedom, }
and the weights are given by
  \begin{equation}
\label{eq_weights_general}
\displaystyle w(\lambda_n) \define
	\displaystyle   h_{a}(\lambda_n)h_{b}^2(\lambda_n),~\forall n \in {\text{supp}}(h_b).
\end{equation}
\end{proposition}
\begin{IEEEproof} 
	The proof appears in Appendix \ref{Performance_Analysis}.
\end{IEEEproof}
While a closed form expression for {$Q$} 
exists only for $M=2$ (pp.  152-153, \cite{Kay_detection}), there are many approximations in the literature that can be used for any $M$ \cite{bodenham2016comparison}.
\section{Smoothness Detector}\label{sub_smooth_detect}
The objective of this study is to investigate the smoothness of signals measured over the vertices of a given graph.
Specifically, we aim to determine if these graph signals can be considered as  the output of a smooth graph filter  with a white Gaussian noise input.
To this end,  we assume that the  measurements obey the model in \eqref{model}, and consider the following hypothesis testing problem: 
\beqna\label{eq:hypothesis}
 \begin{array}{l}   \mathcal{H}_0: \text{the graph filter ${h}(\Lmat)$ is smooth},\\
    \mathcal{H}_1: \text{the graph filter ${h}(\Lmat)$ is not smooth},
    \end{array}
\eeqna
where a smooth graph filter is defined in Definition \ref{smoothdef}, and under $\mathcal{H}_1$ the graph filter, ${h}(\Lmat)$, does not satisfy Definition \ref{smoothdef}.

To tackle the conceptual hypothesis testing problem in \eqref{eq:hypothesis},  we consider two cases. In Subsection \ref{sub;detectors_known_model}, we examine the case of known model parameters and develop the associated \ac{lrt}.  
In Subsection \ref{sec;semi_parametric},  we propose a semi-parametric detector for the case 
where the graph filters and the input signal variance are unknown.
{{Finally, in Subsection \ref{comp_subsec}, we discuss the computational complexity of the methods.}}

\subsection{Detectors Based on a Known Model}\label{sub;detectors_known_model}
Assuming that the measurements $\{\xvec[m]\}_{m=1}^M$ were generated according to the model in \eqref{model}, \eqref{eq:hypothesis} can be written as
 \beqna\label{eq:hypothesis2}
 \begin{array}{l}
    \mathcal{H}_0: \xvec[m]=h_{0}(\Lmat)\yvec[m]
 \\
    \mathcal{H}_1: \xvec[m]=h_{1}(\Lmat)\yvec[m]
    \end{array},~~~
    m=1,\ldots,M,
\eeqna
where $h_{0}(\Lmat)$ and $h_{1}(\Lmat)$  are known  smooth and non-smooth graph filters, respectively.
According to the model described in Subsection \ref{subsec_measurements_model}, $\{\yvec[m]\}_{m=1}^M$  are \ac{iid}  Gaussian random vectors with zero mean and a known covariance matrix, $\sigma^2\Imat$. We assume that under both hypotheses from  \eqref{eq:hypothesis2}
\[\sigma^2\|h_0(\Lmat)\|_2^2={\rm{E}}\left[\xvec^T[m]\xvec[m]\right]=\sigma^2\|h_1(\Lmat)\|_2^2.\]
Therefore, the norm of the filters should be equal, i.e. $||h_{0}(\pmb{\lambda})||_2^2=||h_{1}(\pmb{\lambda})||_2^2$; otherwise, the hypotheses can be distinguished according to the norm of $\{\xvec[m]\}_{m=1}^M$, which does not indicate the signal smoothness. 
{Given that ${\yvec[m]}_{m=1}^M\sim\mathcal{N}(\zerovec,\sigma^2\Imat)$, it follows from \eqref{eq:hypothesis2} that $\xvec[m]|\mathcal{H}_i \sim\mathcal{N}(\zerovec, \sigma^2h_{i}^2(\Lmat))$. The log-likelihood of $\xvec$ under hypothesis $\mathcal{H}_i$ is determined following the derivation 
for the (singular) Gaussian case (see, e.g. p. 73 in \cite{Kayestimation}, p. 276 in \cite{Khatri1968}) as follows:}
\beqna
\label{log_trans}
\log f(\xvec|\mathcal{H}_i)
=\frac{M}{2\sigma^2}\log(|(h_{i}^2(\Lmat))^{\dagger}|_{+})\hspace{1.5cm}\nonumber\\-\frac{1}{2\sigma^2}
\sum\nolimits_{m=1}^M\xvec^T[m](h_{i}^2(\Lmat))^{\dagger}\xvec[m],
\eeqna
$i=0,1$,
where ${\xvec}=[{\xvec}^T[1],\ldots,{\xvec}^T[M]]^T$.

The use of the pseudo-inverse and the pseudo-determinant in \eqref{log_trans} is since $h_{i}(\Lmat)$ may be a singular matrix. For example, for the \ac{gmrf} graph filter in \eqref{GMRF filter} and $\pmb{\mu}=\zerovec$ in \eqref{filter_mu}, we obtain $h_{0}^2(\Lmat)=\Lmat$, 
 which is a singular matrix. 

The hypothesis testing problem outlined in (\ref{eq:hypothesis2}) assumes that $ h_{0}(\cdot)$, $h_{1}(\cdot)$, $\Lmat$,  and the statistics of the input signal are all known. 
This problem is equivalent to testing the structured covariance matrix of Gaussian random vectors, which has been extensively studied in the literature (see, e.g. \cite{Kay_detection,cov1,Ramirez}).

The \ac{lrt}  for this case, after removing constant terms, is 
\beqna\label{LRT}
    \text{LRT}(\xvec)
    =\log\Big(\frac{f(\xvec|\mathcal{H}_1)}{f(\xvec|\mathcal{H}_0)}\Big)\hspace{4cm}
   \nonumber\\ \propto\frac{M}{2\sigma^2} \text{Tr}\left( \Smat_{\xvec}\left((h_{0}^2(\Lmat))^{\dagger}- (h_{1}^2(\Lmat))^{\dagger}\right)\right)    \mathop{\gtrless}^{H_1}_{H_0}\,\gamma,
\eeqna
where  
    $\Smat_{\xvec}\triangleq \frac{1}{M}\sum_{m=1}^M\xvec[m]\xvec^{T}[m]$ is the sample covariance matrix.
The second equality in \eqref{LRT} is obtained by substituting \eqref{log_trans} and using the properties of the trace operator.
 Thus, $\Smat_{\xvec}$ is a sufficient statistic of the \ac{lrt} in the time domain.
   \subsubsection{Interpretation of the LRT} 
 By using the Frobenius norm  in \eqref{LRT},  the detector from \eqref{LRT} can be written as 
  \beqna\label{LRT_rho2}
 \text{LRT}(\xvec)=\frac{M}{2\sigma^2}||\Smat_{\xvec}^{1/2}h_{0}^{\dagger}(\Lmat)||_F^2
 -\frac{M}{2\sigma^2}||\Smat_{\xvec}^{1/2}h_{1}^{\dagger}(\Lmat)||_F^2. 
\eeqna
As the Frobenius norm is a vector norm in the $N\times N$ dimensional space \cite{Horn2012}, the detector specified by  \eqref{LRT_rho2} can be viewed as measuring the energy difference between the projection of $\Smat_{\xvec}$ on $h_{0}^{\dagger}(\Lmat)$ and its projection on  $h_{1}^{\dagger}(\Lmat)$. 
  
 By substituting the \ac{evd} of $h_{0}(\Lmat)$ and $h_{1}(\Lmat)$ in \eqref{LRT}, using the \ac{gft} from \eqref{GFT}, and the trace operator properties, we get  
    \beqna\label{LRT_gft}
 \text{LRT}(\xvec)\hspace{6.5cm}\nonumber\\=\frac{1}{2\sigma^2}
\sum_{n=1}^N\left(((h_{0}^2(\lambda_n))^{\dagger}- (h_{1}^2(\lambda_n))^{\dagger})\sum_{m=1}^M\tilde{\xvec}_n^2[m]  \right).
\eeqna
Thus, the \ac{lrt} is a weighted sum of the squares of the frequency components of the measurements, $\sum_{m=1}^M\tilde{\xvec}_n^2[m]$, weighted by $\left((h_{0}(\lambda_n))^{\dagger}- (h_{1}(\lambda_n))^{\dagger}\right)$. 
If the energy of the non-smooth graph filter, $h_{1}(\cdot)$, is concentrated in the high frequencies and the energy of the smooth graph filter, $h_{0}(\cdot)$, is concentrated in the low frequencies, then 
$\left((h_{0}(\lambda_n))^{\dagger}- (h_{1}(\lambda_n))^{\dagger}\right)$ are positive for large indices $n$, and are negative for small indices. 
Therefore, if the frequency components of the measurements that correspond to the low (high) frequencies are dominant, the weighted sum in \eqref{LRT_gft} will be negative (positive). 
{
In addition, this \ac{lrt} is similar to the non-parametric detector from \cite{preti2019decoupling}, which calculates the local smoothness of each node based on the {\em{ratio}} between the norm of the filtered output of graph high-pass filter and graph \ac{lpf} (instead of subtraction, as in \eqref{LRT_gft}).
However, the detector in \cite{preti2019decoupling}  assesses each node separately and, thus, is not useful for the validation of the global smoothness considered here.}
 \subsubsection{Performance analysis}
 In order to calibrate the threshold $\gamma$ and analyze the detector performance, the \ac{lrt} distribution has to be determined.  
According to Proposition \ref{prop}, 
the detector from \eqref{LRT_gft} under hypothesis $\mathcal{H}_0$ is a weighted sum of 
 independent $\chi$-square random variables with $M$ degrees of freedom. The weights in this case are obtained by substituting the graph filters  $h_{a}(\cdot)=\frac{1}{\sigma^2}(h_{0}^2(\cdot))^{\dagger}-(h_{1}^2(\cdot))^{\dagger})$ and $h_{b}(\cdot)=\sigma h_{0}(\cdot)$ in \eqref{eq_weights_general}, which results in  
 \begin{align}
\label{eq_weights_LRT}
\displaystyle w(\lambda_n) 
=h_{0}^2(\lambda_n)\left((h_{0}^2(\lambda_n))^{\dagger}-(h_{1}^2(\lambda_n))^{\dagger}\right),
\end{align}
$\forall  n \in {\text{supp}}(h_{0})$.
 Substituting \eqref{eq_weights_LRT} in \eqref{eq_pfa_general}, and approximating the inversion of $Q({M,\{w(\lambda_n)\}_{n \in {\text{supp}}(h_b)}},\gamma)$ by that given in e.g. \cite{bodenham2016comparison}, gives the threshold, $\gamma$, for the \ac{lrt}, which keeps the false alarm probability under a predefined level. 

  \subsubsection{Special cases}
If under hypothesis $\mathcal{H}_0$, $h_{0}(\Lmat)$ is one of the smooth graph filters from Box 1 (\eqref{GMRF filter}, \eqref{Tikhonov filter}, and \eqref{diffusion filter}), 
while under $\mathcal{H}_1$, 
$h_{1}(\Lmat)=\Imat$ with $r=1$ (see Fig. \ref{fig_filters}), then 
by substituting $h_{0}^2(\Lmat)$ and $h_{1}^2(\Lmat)$ in \eqref{LRT} we get
\beqna 
\text{LRT}_\text{GMRF}(\xvec)=\frac{M}{2\sigma^2} \text{Tr}\left(\Smat_{\xvec}\left(\beta_{GMRF}^{-2}\Lmat-\Imat\right)\right)\label{eq:LRT_GMRF},\hspace{1.5cm}\\
\text{LRT}_\text{Tikh.}(\xvec)=\frac{M}{2\sigma^2} \text{Tr}\left(\Smat_{\xvec}\left(\beta_{Tikh.}^{-2}(\Imat+\alpha\Lmat)^{2}-\Imat\right)\right)\label{eq:LRT_tikhonov},\hspace{0.5cm}\\
\text{LRT}_\text{Diff.}(\xvec)=\frac{M}{2\sigma^2} \text{Tr}\left(\Smat_{\xvec}\left(\beta_{Diff.}^{-2}\exp{(2\tau\Lmat)}-\Imat\right)\right).\hspace{0.25cm}\label{eq:LRT_diffusion}
\eeqna 
Since the Laplacian is a \ac{gso}, $\Lmat^k$ aggregates each node with its $k$ neighboring nodes, i.e.
\begin{equation}
    [\Lmat\avec]_i=\sum_{j\in\mathcal{N}_i}w_{i,j}(a_i-a_j),
\end{equation}
where ${\mathcal{N}}_i= \{k \in {\nu} : (i,k) \in \xi\}$ is the 1-hop neighborhood of node $i\in {\nu}$,
and, in a recursive manner, higher powers perform a weighted sum of the differences from the lower orders:
\begin{equation}\label{eq_laplacian_k_power_shift}
    [\Lmat^k\avec]_i=\sum_{j\in\mathcal{N}_i}w_{i,j}([\Lmat^{k-1}\avec]_i-[\Lmat^{k-1}\avec]_j).
\end{equation}
In particular, for $(h_{0}(\Lmat))^{\dagger}=\sum_{k=0}^Kh_k\Lmat^k$, then 
\begin{equation}\label{eq_poly}
    \sum_{m=1}^M\xvec^T[m](h_{0}(\Lmat))^{\dagger}\xvec[m]=\sum_{m=1}^M\xvec^T[m]\left(\sum_{k=0}^Kh_k\Lmat^k\xvec[m]\right).
\end{equation}
Using \eqref{eq_laplacian_k_power_shift}, it can be seen that the r.h.s. of \eqref{eq_poly} is a weighted sum of the differences of the graph signal in each node from its neighbors up to order $K$.
The weights of the filter $\{h_k\}_{k=0}^K$ control the importance of the similarity of the node to its $k$-hop neighbors, $k=1,\ldots,K$. 
Thus, the closer the signal values of a node and its neighbors are, the smaller $\sum_{m=1}^M\xvec^T[m](h_{0}(\Lmat))^{\dagger}\xvec[m]$ will be. Since $(h_{1}(\Lmat))^{\dagger}=\Imat$, this term is compared to the value of the signal on each individual node without the aggregation of its neighbors, resulting in a smaller detector value, since most of the signal values on the different nodes are similar (i.e. the output graph signal is smooth). As a result, 
the GMRF-filter-based \ac{lrt} from  \eqref{eq:LRT_GMRF} tests similarity with up-to-first-order neighbors, the Tikhnov-filter-based detector from \eqref{eq:LRT_tikhonov} tests similarity up-to-second-order neighbors, and the diffusion-filter-based detector from \eqref{eq:LRT_diffusion} 
aggregates all nodes in the graph with weights in descending order based on their distances (can be seen from the Taylor expansion $\exp{(2\tau\Lmat)}=\sum_{k=0}^{\infty}\frac{1}{k!}\Lmat^k$, p. 351, \cite{Horn2012}). Thus, the \ac{lrt}s with filters from Box 1 represent smoothness measures that rely on aggregating neighboring values of varying orders, where the order depends on the specific filter. 

\subsubsection{{Relation with Definition \ref{def_LPF}}}
{If one uses Definition \ref{def_LPF} instead of Definition \ref{smoothdef} for the hypothesis testing problem in \eqref{eq:hypothesis2}, the \ac{lrt} will be the same as in \eqref{LRT} with the requirement that $h_0(\Lmat)$ and $h_1(\Lmat)$ satisfy Definition \ref{def_LPF}. Thus, 
under the null hypothesis $\mathcal{H}_0$, graph filters that satisfy the \ac{lpf} conditions in Definition \ref{def_LPF} and the conditions in Claim \ref{claim_equiv_to_lpf} are also smooth graph filters. However, under the alternative hypothesis $\mathcal{H}_1$, defining a non-\ac{lpf} graph filter using Definition \ref{def_LPF} is not straightforward: 
A filter that is not an \ac{lpf} from order $K$ may be a $K+1$ \ac{lpf}.
In contrast, since Definition \ref{smoothdef} considers all filter coefficients, if a filter does not satisfy  Definition  \ref{smoothdef}, there is a greater probability that it is not smooth. Thus, using Definition \ref{smoothdef} is more suitable for testing smoothness.}
\subsubsection{Discussion}
 The advantages of the \ac{lrt} in \eqref{LRT} are its simplicity, the fact that there is no need to calculate the \ac{evd} of $\Lmat$, and the ability to calibrate the threshold to bound its probability of false alarm.
 However, it is based on assuming known parameters: $\sigma^2$, $h_{0}(\Lmat)$, and $h_{1}(\Lmat)$. 
 Thus, if the measurement model is unknown, the detector performance might be degraded due to a model mismatch.

\subsection{Semi-parametric Detector}\label{sec;semi_parametric}
In this subsection, 
 we develop a semi-parametric approach that is not based on known parameters. In particular, 
we suggest deriving the \ac{ml} estimator of ${h}^2(\Lmat)$ and $\sigma^2$ from the model in \eqref{model}, and then checking if the estimated filter is smooth according to Definition \ref{smoothdef}. To restrict ${h}(\Lmat)$ to be a graph filter, we use \eqref{graph_filter}  and replace ${h}(\Lmat)$ with $\Vmat\text{diag}({h}(\pmb{\lambda}))\Vmat^{-1}$ in the log-likelihood in \eqref{log_trans}. Accordingly, 
the \ac{ml} estimators of ${h}(\pmb{\lambda})$ and $\sigma^2$ are
\beqna
\label{log_semi_ML}
\{\hat{h}(\pmb{\lambda}),\hat{\sigma}^2\}
\hspace{6.25cm}\nonumber\\
=\arg\max_{{h}(\pmb{\lambda})\in\mathbb{R}^{N},\sigma^2\in\mathbb{R}_{+}}\frac{M}{2}\log(|\Vmat \text{diag}(\sigma^2{h}^2(\pmb{\lambda}))^{\dagger}\Vmat^{-1}|_{+})
\nonumber\\-\frac{1}{2}
\sum_{m=1}^M\xvec^T[m](\Vmat \text{diag}(\sigma^2{h}^{2}(\pmb{\lambda}))^{\dagger})\Vmat^{-1}\xvec[m].
\eeqna
Now we define 
\[\mathcal{D}\define\{ n|n\in\{1,\ldots,N\},~\sum_{m=1}^M\tilde{x}_n^2[m]\not=0\}.\]
By using $\mathcal{D}$, we can get rid of the pseudo-determinant and pseudo-inverse terms.
In particular, 
by using  $\mathcal{D}$, the \ac{gft} definition in \eqref{GFT}, and the fact that  the unitary matrix determinant is $1$, we can write \eqref{log_semi_ML} in the graph frequency domain as
\beqna
\label{log_semi_gft}
\{\hat{\sigma}^2,\{\hat{h}(\lambda_n)\}_{n=1}^N\}\hspace{5cm}\nonumber\\=\arg\max_{\{{h}(\lambda_n)\}\in\mathbb{R},\sigma^2\in\mathbb{R}_{+}}
\frac{M}{2}\sum_{n\in\mathcal{D}}\log((\sigma^2h^2(\lambda_n))^{-1})\nonumber\\-\frac{1}{2}
\sum_{n\in\mathcal{D}} (\sigma^2h^2(\lambda_n))^{-1}\sum_{m=1}^M\tilde{\xvec}_n^2[m].\hspace{1.8cm}
\eeqna

Similar to the explanation in Subsection \ref{subsec_measurements_model} (before Claim \ref{prop_mu}),  $\sigma^2$ and ${h}(\pmb{\lambda})$ cannot be uniquely determined from \eqref{log_semi_ML} without further information. Since, according to Definition \ref{smoothdef}, scaling does not affect the smoothness of the estimated graph filter, we derive the \ac{ml} of the multiplication  $\sigma^2h^2(\lambda_n)$.
By equating the derivative of the log-likelihood function from \eqref{log_semi_gft} w.r.t. $\sigma^2 {h}^2(\lambda_{n})$,  $ n\in\mathcal{D}$, to zero, the \ac{ml} estimators satisfy
\begin{equation}\label{opt_A}
\hat{\sigma}^2\hat{h}^2(\lambda_n) =
    \frac{1}{M}\sum_{m=1}^M\tilde{\xvec}_{n}^2[m],~~\forall n\in\mathcal{D}.
\end{equation}

The solution in \eqref{opt_A} holds under the assumption  that 
all the eigenvalues of $\Lmat$ are distinct.
Otherwise, 
the graph filter should satisfy the constraint in \eqref{constraint1}.
As a result,
the \ac{ml} estimator in \eqref{opt_A} should be replaced by the constrained \ac{ml} estimator. In  Appendix \ref{appendix_filter_coeff} it is shown that the constrained \ac{ml} is given by
\begin{equation}\label{opt_A2}
\hat{\sigma}^2\hat{h}^2(\lambda_n) = \frac{1}{|{\mathcal{S}}_{n}|}\sum_{j\in {\mathcal{S}}_n} \frac{1}{M}\sum_{m=1}^M\tilde{\xvec}_j^2[m]~~\forall n\in\mathcal{D}, 
\end{equation}
where 
${\mathcal{S}}_n=\{j=1,\ldots,N|\lambda_j=\lambda_n\}$ is the set of the indices of all eigenvalues that are equal to $\lambda_n$.
It can be seen that \eqref{opt_A2} implies that 
$\hat{\sigma}^2\hat{h}^2(\lambda_m) = \hat{\sigma}^2\hat{h}^2(\lambda_k)$, $\forall \lambda_m=\lambda_k$. 
Thus, \eqref{opt_A2} keeps the amplitudes of the graph frequency response of the same eigenvalues identical, s.t. $\hat{h}(\lambda)$ is a valid graph filter.

Now, we should check if the estimated graph filter in \eqref{opt_A2}, $\hat{h}(\lambda)$, which represents the graphical process that generates the data, is a smooth graph filter by Definition \ref{smoothdef}.
By substituting  \eqref{opt_A2} in the smoothness condition from \eqref{eq_SGF_gft}, we get the condition
 \beqna\label{eq_semi_detector}
      \hat{r}=\lambda_{\rm{avg}}^{-1}
      \frac{\sum_{n\in\mathcal{D}}{\sum_{m=1}^M {{\lambda_n}}\tilde{\xvec}_n^2[m]}}{\sum_{n\in\mathcal{D}}\sum_{m=1}^M\tilde{\xvec}_n^2[m]}
<1, \eeqna {where the sums in the numerator are weighted by the eigenvalues, $\lambda_n$.}
Alternatively, 
by using the \ac{gft} and \ac{tv} definitions from \eqref{GFT}-\eqref{eq:Dirichlet energy}, the detector from \eqref{eq_semi_detector} can be rewritten as
  \beqna\label{eq_semi_detector2}
\hat{r}=\lambda_{\rm{avg}}^{-1}\frac{\sum_{m=1}^M\xvec^T[m]\Lmat\xvec[m]}{\sum_{m=1}^M||\xvec[m]||^2}
<1. \eeqna

\subsubsection{Discussion} The detector in \eqref{eq_semi_detector2} can be interpreted as a normalized version of the empirical expectation of the \ac{tv}.
Accordingly, in the absence of knowledge about the filter that generated the measurements, the detection approach is based on evaluating the smoothness empirically based on   \eqref{eq:Dirichlet energy} and normalizing it by the sample variance of the data to mitigate the impact of scaling on the result.
In addition, from the expression in the \ac{gft} domain in \eqref{eq_semi_detector}, it can be seen that we compare the weighted average of $\pmb{\lambda}$ (weighted by the frequency components of the signal) with the average of $\pmb{\lambda}$ ($\lambda_{\rm{avg}}$). Finally, 
as demonstrated in the simulations, 
in practice, the proposed semi-parametric detector achieves the probability of detection of the \ac{lrt} (with known graph filters), which is an upper bound on the detection performance, and, thus, is optimal for the tested scenarios. 

  \subsubsection{Special cases} In the special case where  $\sum_{m=1}^M\tilde{\xvec}_n[m]=\sum_{m=1}^M\tilde{\xvec}_k[m]$ for all $n\not=k$, we obtain $\hat{r}=1$, and the signal is not identified as smooth. In this case, the energy of the signals is uniformly distributed across all graph frequencies, and thus, the signal has a large \ac{tv}. In contrast,if the energy of the graph signals tends to lie in the low frequencies, then the weighted average in $\hat{r}$ will be lower than the regular average, and the filter will be considered smooth. 
  
\subsubsection{Setting the threshold} In the general case, there is no analytical way 
 to tune the threshold of the semi-parametric detector.
 As an ad-hoc method, we  can use the special case of a partially known measurements model, given by the hypothesis testing in \eqref{eq:hypothesis} with known $h_{1}^2(\Lmat)=\Imat$ and $h_{0}^2(\Lmat)=\Lmat^{\dagger}$, and unknown $\sigma^2$ for both hypotheses. In this case, 
the expression in the r.h.s. of \eqref{eq_semi_detector2} can be shown to be the \ac{glrt}, where $\sigma^2$ should be  replaced by its \ac{ml} estimator under the different hypotheses (see proof in Appendix \ref{sub;detectors_partially_known_model}). In this case, 
under hypothesis $\mathcal{H}_0$, $\xvec[m]\sim\mathcal{N}(\zerovec, \sigma^2\Lmat^{\dagger})$, and
the probability of false alarm is
\beqna\label{eq_pfa_semi}
 \Pr\left( \hat{r} >\gamma\right)= \Pr\left(\frac{\sum_{m=1}^M \xvec^T[m] \Lmat\xvec[m]}{\sum_{m=1}^M \xvec^T[m] \xvec[m]}>\gamma\right)
   \nonumber\\=   \Pr\left(\sum_{m=1}^M \xvec^T[m] (\Lmat-\gamma\Imat)\xvec[m]>0\right).
\eeqna
Thus, for a given $\gamma$, we can use Proposition \ref{prop} where $h_{a}(\lambda_n)=\lambda_n-\gamma$, $h_{b}(\lambda_n)=\sqrt{\lambda_n^{\dagger}}$, and find the false alarm probability:
\beqna\label{eq_pfa_semi2}
 \Pr\left( \hat{r} >\gamma\right)=1-Q\left({M,\Big\{1-\frac{\gamma}{\lambda_n}\Big\}_{n \in \{2,\ldots,N\}}},\gamma\right).
 \eeqna
 Inverting \eqref{eq_pfa_semi2} gives the threshold for the semi-parametric detector, such that the false alarm probability does
not exceed a predefined level.
Finally, Algorithm \ref{alg2} outlines the semi-parametric method for determining the smoothness of the data.
\begin{algorithm}[hbt]
\textbf{Input:}
\begin{itemize}
\item Laplacian matrix, $\Lmat$.
\item Measurements $\{\xvec[m]\}_{m=1}^M$.
\end{itemize}
\textbf{Output:} A binary decision indicating whether the data is smooth or not.
\begin{algorithmic}[1]
\STATE Calculate  $\hat{\sigma}^2\hat{h}^2(\pmb{\lambda})$ from 
\eqref{opt_A2}.
\STATE Substitute $\hat{\sigma}^2\hat{h}^2(\pmb{\lambda})$ into \eqref{eq_SGF_gft} to calculate $\hat{r}$.
\STATE Utilize \eqref{eq_pfa_semi2} to calculate the threshold value, $\gamma$, corresponding to the desired level of probability of false alarm.
\STATE\textbf{if} $\hat{r}<\gamma$:
 \textbf{Return} ``The data is smooth."
\STATE\textbf{else:}
 \textbf{Return} ``The data is not smooth."
\end{algorithmic}
\caption{Semi-parametric smoothness detector}
\label{alg2}
\end{algorithm}

\subsection{{Computational Complexity}}
\label{comp_subsec}
{
In the following, we discuss the computational complexity of all the methods compared in Section \ref{sec;sim}.
The computational complexity of the semi-parametric and the naive \ac{tv} detectors is $O(N^2 M)$, due to the necessity of computing $M$ times the quadratic form in \eqref{eq_semi_detector2} ($N^2$ multiplications) to calculate the mean across a dataset of size $M$. 
For the first-order \ac{lpf}, computing the estimated covariance matrix (see (6) in \cite{He2012Hoi-To}) is $O(N^2M)$, and finding the minimal eigenvector 
of the estimated covariance has almost linear complexity \cite{LOBPCG}, as the estimated covariance is a positive semi-definite matrix. 
The computational complexity of the \acp{lrt}, the Blind Simple Matched Subspace Detector (BSMSD), and the \ac{lpf} are $O(N^3+N^2M)$, due to their preprocessing steps: The \ac{lrt}s require computing the inverse of the $N$-dimensional matrices in \eqref{LRT}, and the BSMSD and the \ac{lpf} require the \ac{evd} calculation of $\Lmat$,  with of complexity $O(N^3)$ \cite{6969512}. Then, adding the calculation of the quadratic forms in \eqref{LRT}, in (17) from \cite{Isufi2018Leus}, and in \eqref{opt_A}-\eqref{opt_A2}, respectively, sums up to $O(N^3+N^2M)$. Hence, it can be seen that the computational complexity of our method is comparable to the other methods.
}
\section{Simulations}\label{sec;sim}
In this section, the performance of the proposed \ac{lrt}s from \eqref{eq:LRT_GMRF}-\eqref{eq:LRT_diffusion} and the semi-parametric detector from \eqref{eq_semi_detector} are investigated and compared with existing methods -    for synthetic data,  for electrical networks data, {{and for
\ac{wsn} Data}}
in Subsections \ref{subsec;synthtic_data_sim},  \ref{subsec;power_sim}, {{and \ref{subsec;wsn_sim},}} respectively.
In all simulations, the performance is evaluated using  $10,000$ Monte-Carlo simulations.  We used $\alpha=0.2$ and $\tau=0.1$
for the Tikhonov Regularization \ac{lrt}  from \eqref{eq:LRT_tikhonov}  and for the Diffusion Kernel \ac{lrt} 
  from \eqref{eq:LRT_diffusion}, respectively.
  
The methods for comparison in this section are:
    \begin{enumerate}
        \item The first-order \ac{lpf} detector \cite{He2012Hoi-To}  (denoted as $1$st-order LPF), which assumes a  measurement model similar to \eqref{model} and aims to determine if the graph filter 
        is a first-order \ac{lpf}, without using the
graph topology. 
Thus, it requires an extensive dataset to estimate the data covariance matrix. Since this detector is not in the form of a test statistic comparison against a threshold value, its probability of false alarm is constant. 
        \item BSMSD  (Eq. (17), \cite{Isufi2018Leus})  with a cutoff frequency of {$\frac{N}{2}$}, developed for detecting topology changes.
       \item  The naive \ac{tv} detector (denoted as TV), $\sum_{m=1}^M\xvec[m]^T\Lmat\xvec[m]\,\mathop{\gtrless}^{H_1}_{H_0}\,\gamma$, which is based on \eqref{eq:Dirichlet energy}.
      \item {The \ac{lpf}-based semi-parametric detector (denoted by \ac{lpf}), checks if the estimated filter coefficients satisfy Definition \ref{def_LPF}. This detector is obtained
      by multiplying 
       the numerator and the denominator of \eqref{eta_k_def} by $\sigma$ and substituting the estimated scaled filter coefficients $\vert\sigma\hat{h}(\lambda_n)\vert$, from either \eqref{opt_A} or \eqref{opt_A2}, as follows:
\begin{equation}
\label{eta_k_def_est_lpf}
\hat{\eta}_{k}= \frac{\max\{\vert\sigma\hat{h}({\lambda}_{k+1})\vert,\ldots,\vert\sigma\hat{h}({\lambda}_{N})\vert\}}{\min\{\vert\sigma\hat{h}({\lambda}_{1})\vert,\ldots,\vert\sigma\hat{h}({\lambda}_{k})\vert\}}\,\mathop{\gtrless}^{H_1}_{H_0}\,\gamma.
\end{equation}
Here we set $k=\frac{N}{2}$ by trial and error.
}
        \end{enumerate}

\begin{figure*}[hbt]
\captionsetup[subfigure]{labelformat=empty}
     \centering
     \begin{subfigure}[b]{0.3\textwidth}
         \centering
         \includegraphics[width=\textwidth]{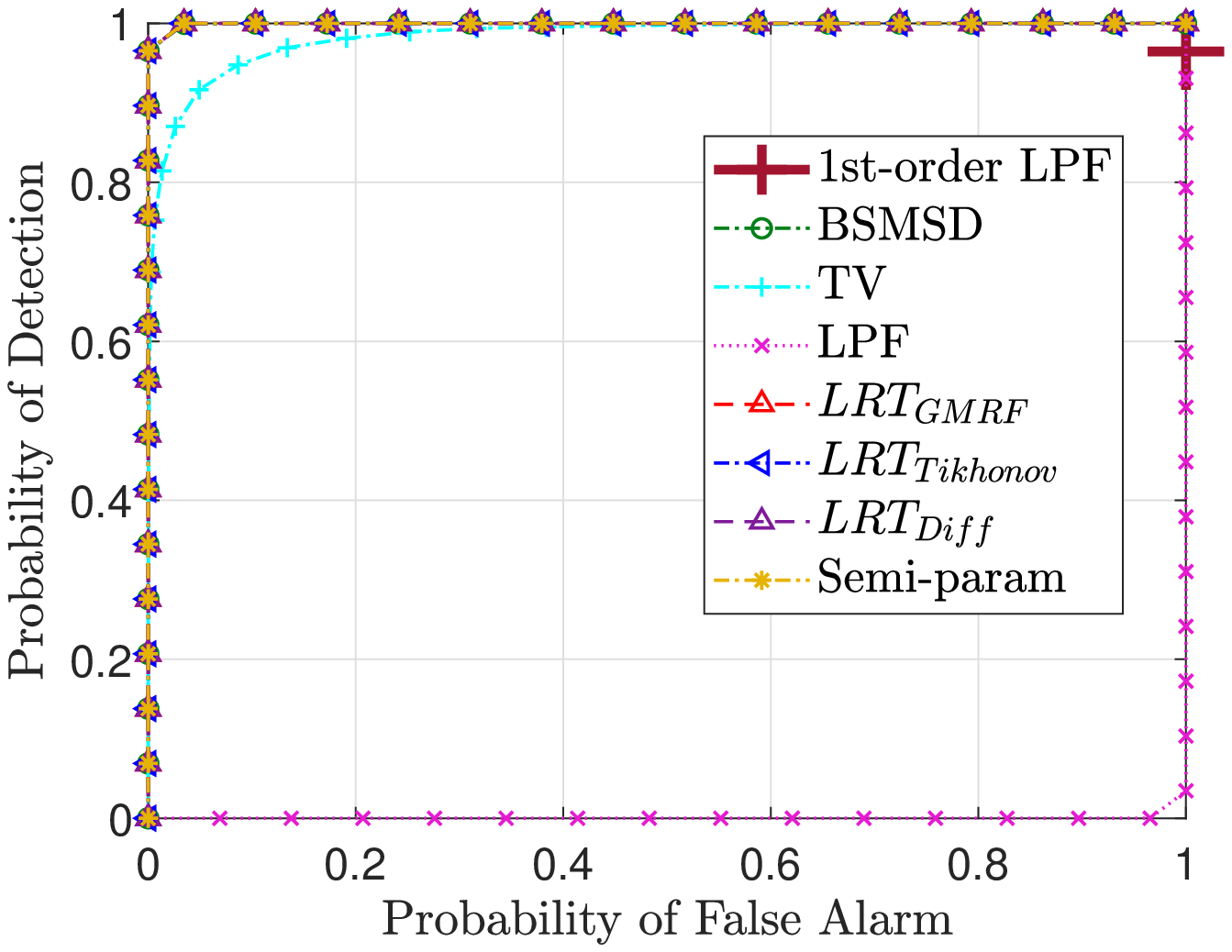}
         \caption{(a)}
         \label{case1}
     \end{subfigure}
     \hfill
     \begin{subfigure}[b]{0.3\textwidth}
         \centering
         \includegraphics[width=\textwidth]{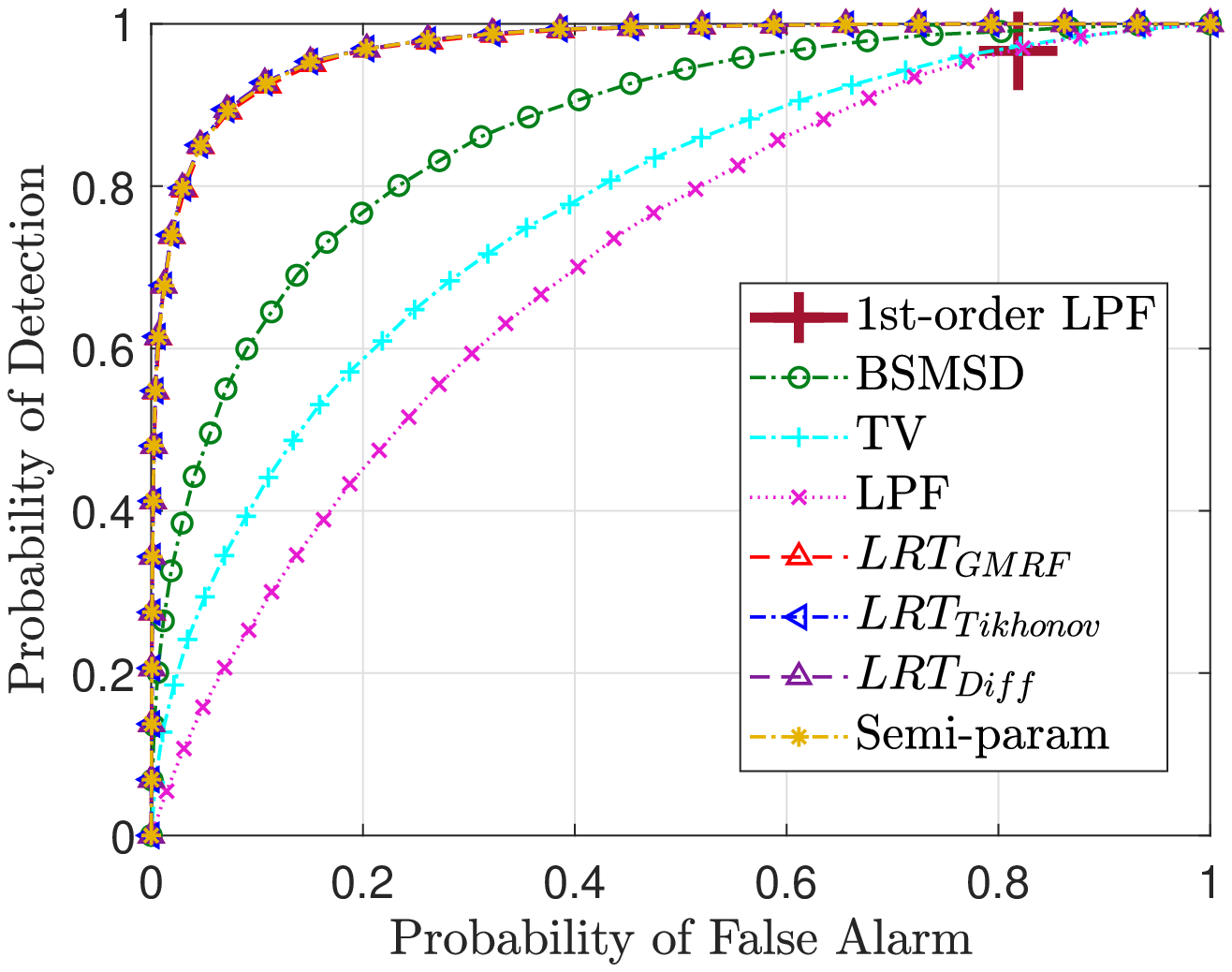}
         \caption{(b)}
         \label{case2}
     \end{subfigure}
     \hfill
     \begin{subfigure}[b]{0.3\textwidth}
         \centering
         \includegraphics[width=\textwidth]{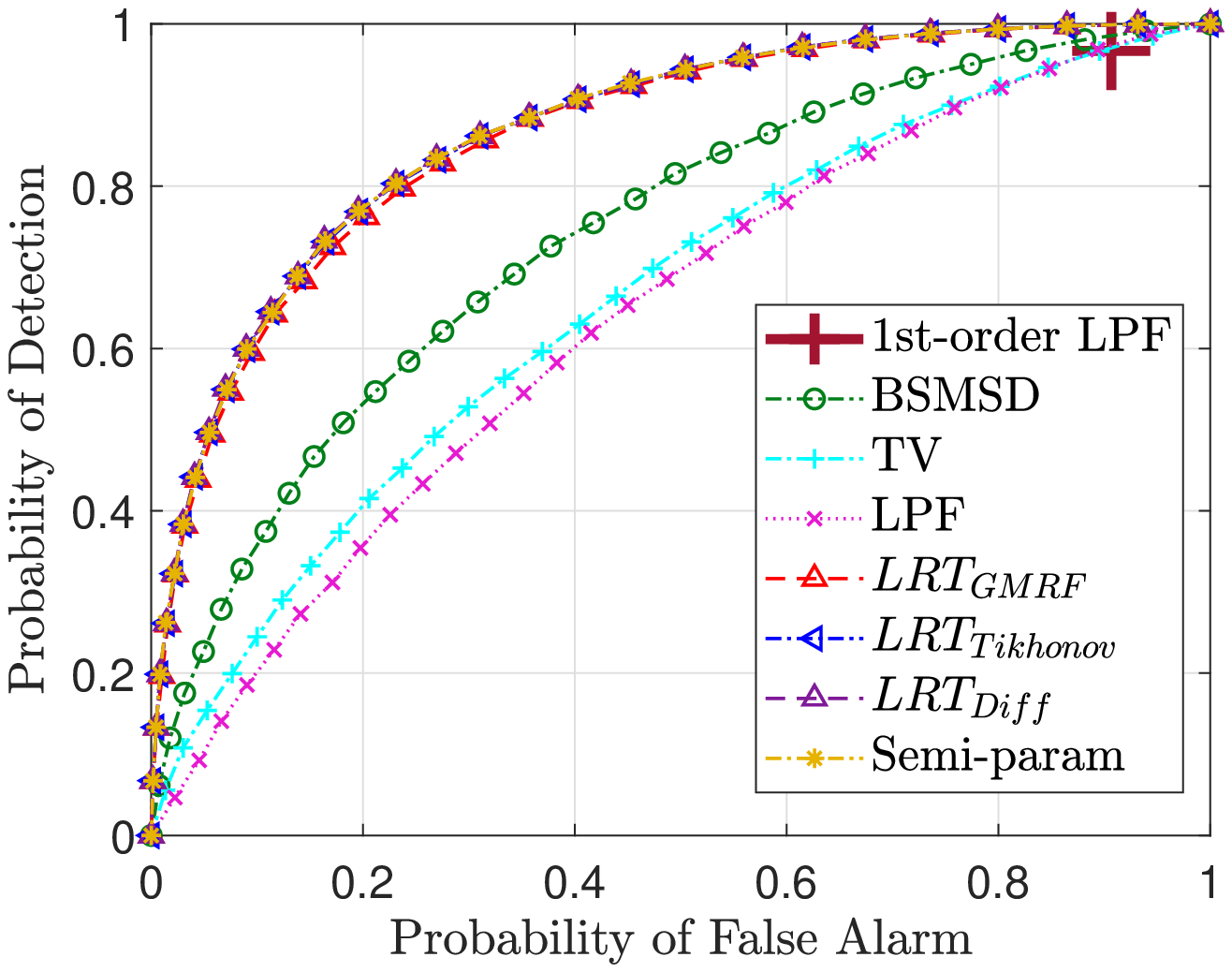}
         \caption{(c)}
         \label{case3}
     \end{subfigure}
        \caption{The \ac{roc} of the {different detectors,} 
        where the graph filters that generated the data are defined in Table \ref{tabSmooth}: (a) $h_{\text{GMRF}}$; (b) $h_{\text{Tikh}}$, $\alpha=0.2$; and (3) $h_\text{Diff}$, $\tau= 0.1$.}
        \label{fig1}
\end{figure*}
\subsection{Synthetic Data}\label{subsec;synthtic_data_sim}
In this subsection, we use a synthetic graph of {$N=30$} vertices whose edges are determined based on Euclidean distances between vertices.
We generated the coordinates of the vertices uniformly
at random in the unit square, and computed the edge
weights with a Gaussian radial basis function:
\begin{equation}\label{rbf_edges}
    \Wmat_{i,j}=\exp\left(-\frac{d(i,j)^2}{2\sigma^2}\right),
\end{equation}
where $d(i,j)$ is the distance between the vertices
and $\sigma=0.5$ is the kernel width. The edges whose weights were smaller than $0.55$ were removed.
Then, unless otherwise stated, we generated $M=30$ \ac{iid} signals $\{\xvec_i\}_{i=1}^{M}$ according to the model in \eqref{model}, where $ h(\Lmat)$ is a smooth graph filter under $ \mathcal{H}_0$ and $ h(\Lmat)=\Imat$  under $ \mathcal{H}_1$. An additive white Gaussian noise was added to the measurements under both hypotheses with covariance $0.1^2\Imat$ to demonstrate the robustness to errors.

\subsubsection{Different smooth generative models}
Figures \ref{fig1}.a-\ref{fig1}.c present the \ac{roc} curves of the three \ac{lrt}s (with the Tikhonov Regularization, the \ac{gmrf}, and the Diffusion Kernel filters), the semi-parametric, the first-order \ac{lpf}, the BSMSD, the naive \ac{tv}, {and the \ac{lpf}} detectors, where the graph filters that were used to generate the data under $\mathcal{H}_0$ appear in  Table \ref{tabSmooth} and  Box 1; they were all normalized such that $\text{Tr}(h^2(\Lmat))=N$.
\begin{table}[hbt]
\begin{center}
\begin{tabular}{|c|c|c|c|}
\hline
 \textbf{Figure} & \textbf{\ref{fig1}.a}& \textbf{\ref{fig1}.b}& \textbf{\ref{fig1}.c} \\
\hline
$h_0(\Lmat)$  &$h_{\text{GMRF}}$
&$h_{\text{Tikh}}$, $\alpha=0.2$
&$h_\text{Diff}$, $\tau= 0.1$ \\
\hline
\end{tabular}
\caption{Graph filter that was used to create the data.} 
\label{tabSmooth}
\end{center}
\end{table}

In Fig. \ref{fig1}.a, {which focuses on the commonly-used \ac{gmrf} measurement model,} the probability of detection is close to 1 for all detectors {except the \ac{lpf} detector}, even for small probabilities of false alarm, since in this case, the signals are significantly smoother under hypothesis $\mathcal{H}_0$ than under $\mathcal{H}_1$,  and since for the \ac{gmrf}, the first graph frequency component, $h_{\text{GMRF}}(\lambda_1)$, is zero, which emphasizes its difference from the all-pass filter under $\mathcal{H}_1$. {In contrast, the \ac{lpf} detector from \eqref{eta_k_def_est_lpf} is sensitive to zeros in the low frequencies, which makes it not useful as a smoothness  detector under the \ac{gmrf} model.}

In  Figs. {\ref{fig1}.b}-\ref{fig1}.c  it can be seen that the proposed detectors 
outperform all the other methods in terms of
 probability of detection for any false alarm probability.
In these figures, the first-order \ac{lpf} has a high probability of false alarm, since it highly depends on the quality of the estimation of the covariance matrix, which is inaccurate for a small number of samples ($M=30$).
 In addition, the superiority of our methods compared to the BSMSD {and the \ac{lpf}} detector can be explained by the fact that the proposed detectors employ a weighted average of the filtered graph frequency components, with greater weight given to higher graph frequency components. In contrast, the BSMSD is an ideal high-pass filter that disregards all frequency components before the cutoff frequency (here $\lambda_{15}$), {and the \ac{lpf} detector is based on minimum and maximum operations, which are sensitive to noise}. 
The naive \ac{tv} detector achieves the lowest probability of detection in all tested scenarios, since it lacks a normalization or subtracted term representing the non-smooth hypothesis, in contrast to the proposed detectors.
%
Finally, it should be noted that in Fig. \ref{fig1}.a, the \ac{lrt} with the \ac{gmrf} graph filter from \eqref{eq:LRT_GMRF} follows the true distribution of the data under both hypotheses, and can be regarded as an upper bound on the \ac{lrt} performance. Similarly, the \ac{lrt} with the Tikhonov Regularization graph filter from \eqref{eq:LRT_tikhonov}, and the \ac{lrt} with Diffusion Kernel graph filter from \eqref{eq:LRT_diffusion}, can be regarded as upper bounds on the \ac{lrt} performance in Fig. \ref{fig1}.b and Fig. \ref{fig1}.c, respectively. In all of these figures, the proposed detectors achieve the upper bounds on the probability of detection.
\subsubsection{Robustness to the dataset size}
Figure \ref{fig:pd_vs_num_samples} displays the probability of detection versus the dataset size, $M$, for $h_0(\Lmat)=\beta(\Imat+0.2\Lmat)^{-1}$ and $h_1(\Lmat)=\Imat$ (same scenario as in Fig. \ref{fig1}.b). 
The false alarm probability is set to $0.001$ for all detectors, except for the first-order \ac{lpf} detector, where it is not applicable.
It can be seen that the detection probability increases as $M$ increases. The detection probability of the proposed detectors is significantly higher than those of the BSMSD, the naive \ac{tv}, {and the \ac{lpf} detectors}. 

\subsubsection{Sensitivity to the smoothness coefficient $r$}
Figure \ref{fig:pd_vs_r} demonstrates the probability of detection versus $r$ from \eqref{SGF_normalized} under $\mathcal{H}_0$ for the probability of false alarm of $0.001$. The data is generated according to the  model in \eqref{model}, where $ h_0(\Lmat)=\beta(\Imat+\alpha\Lmat)^{-1}$   and $ h_1(\Lmat)=\Imat$. The value of $r$ is determined by controlling the value of $\alpha$, where $r=\sum_{n=1}^N\lambda_n(1+\alpha\lambda_n)^{-2}/(\lambda_{\text{avg}}\sum_{n=1}^N(1+\alpha\lambda_n)^{-2})$.
For a fair comparison, the false alarm probability is set to $0.001$ for all detectors, except the first-order \ac{lpf} detector, which cannot be adjusted. 
It can be seen in Fig. \ref{fig:pd_vs_r}  that the probability of detection of all the detectors decreases as $r$ increases. This is because as $r$ increases, the data becomes less smooth according to Definition \ref{smoothdef} (in particular, for $r = 1$, the filter is not smooth), resulting in a higher threshold required to attain the desired probability of false alarm, which reduces the probability of detection. Additionally, the probability of detection for the BSMSD and the \ac{tv} detectors is significantly lower than that of the other detectors for $0.8<r<1$, which highlights their reduced sensitivity to weakly smooth filters.

\subsubsection{Robustness to scaling}
\begin{figure*}[hbt]
\captionsetup[subfigure]{labelformat=empty}
     \centering
     \begin{subfigure}[b]{0.28\textwidth}
         \centering
         \includegraphics[width=1.08\textwidth]{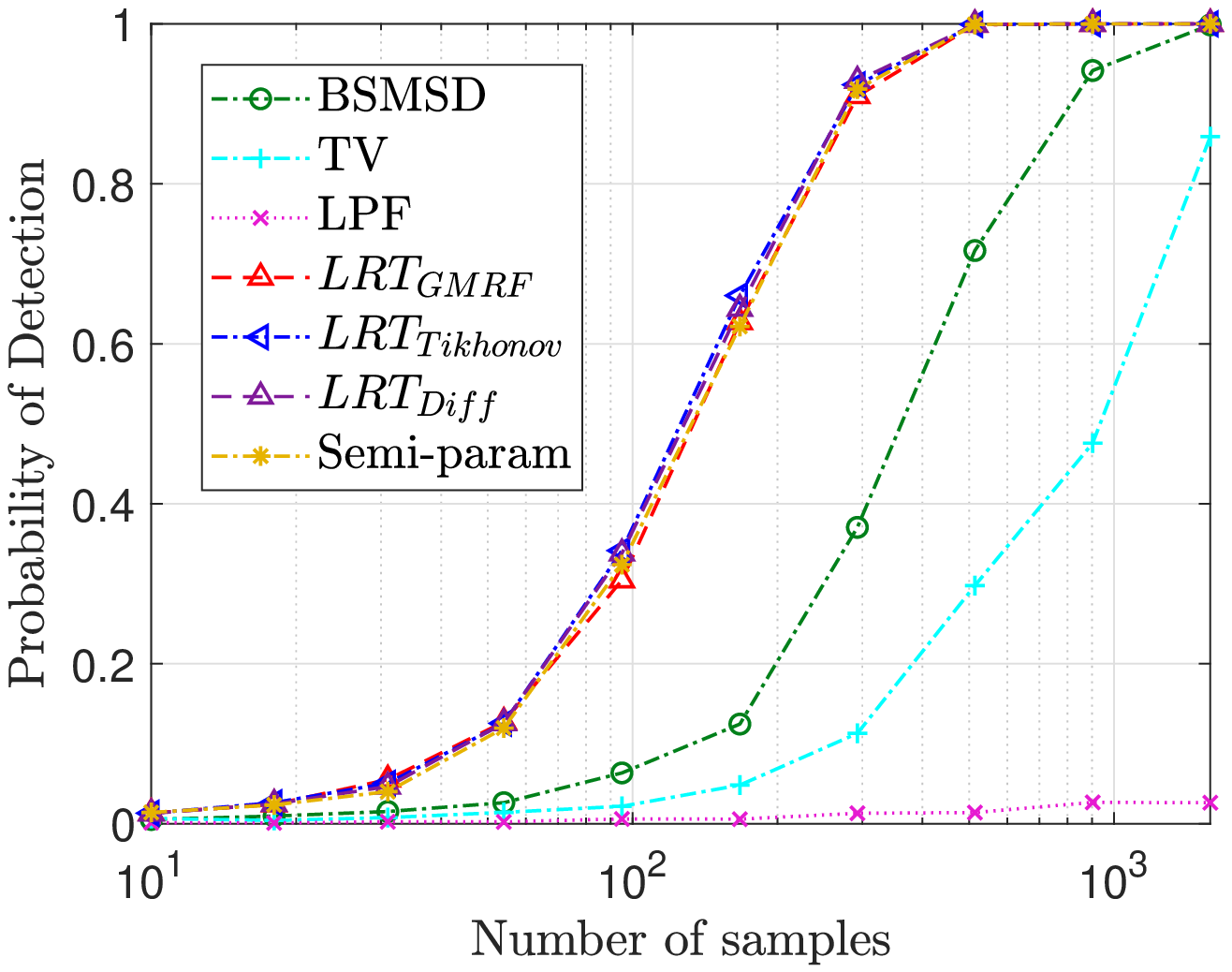}
         \caption{(a)}
\label{fig:pd_vs_num_samples}
     \end{subfigure}
     \hfill
     \begin{subfigure}[b]{0.35\textwidth}
         \centering
         \includegraphics[width=1.08\textwidth]{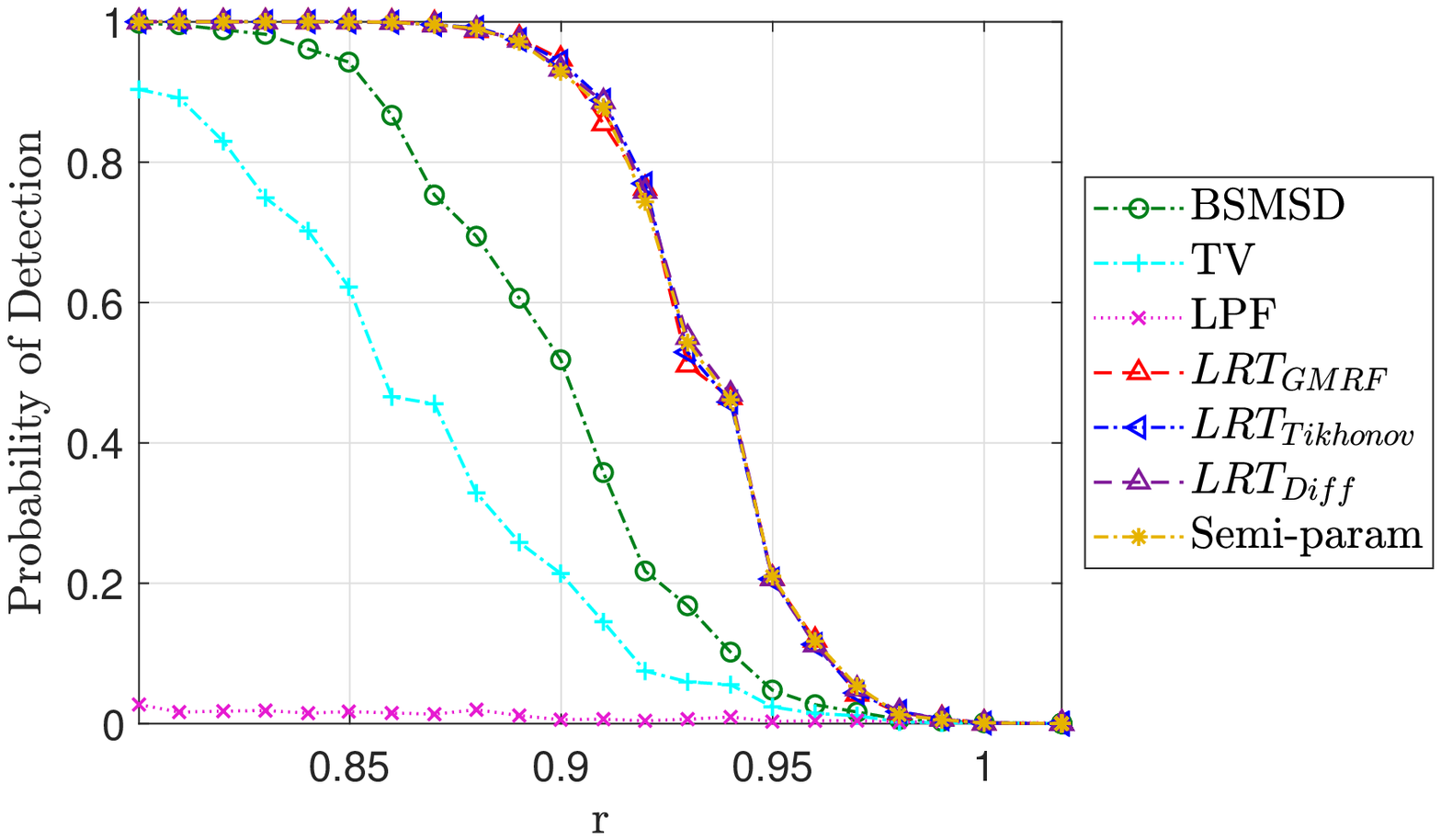}
         \caption{(b)}
\label{fig:pd_vs_r}	
     \end{subfigure}
     \hfill
     \begin{subfigure}[b]{0.35\textwidth}
         \centering
         \includegraphics[width=1.08\textwidth]{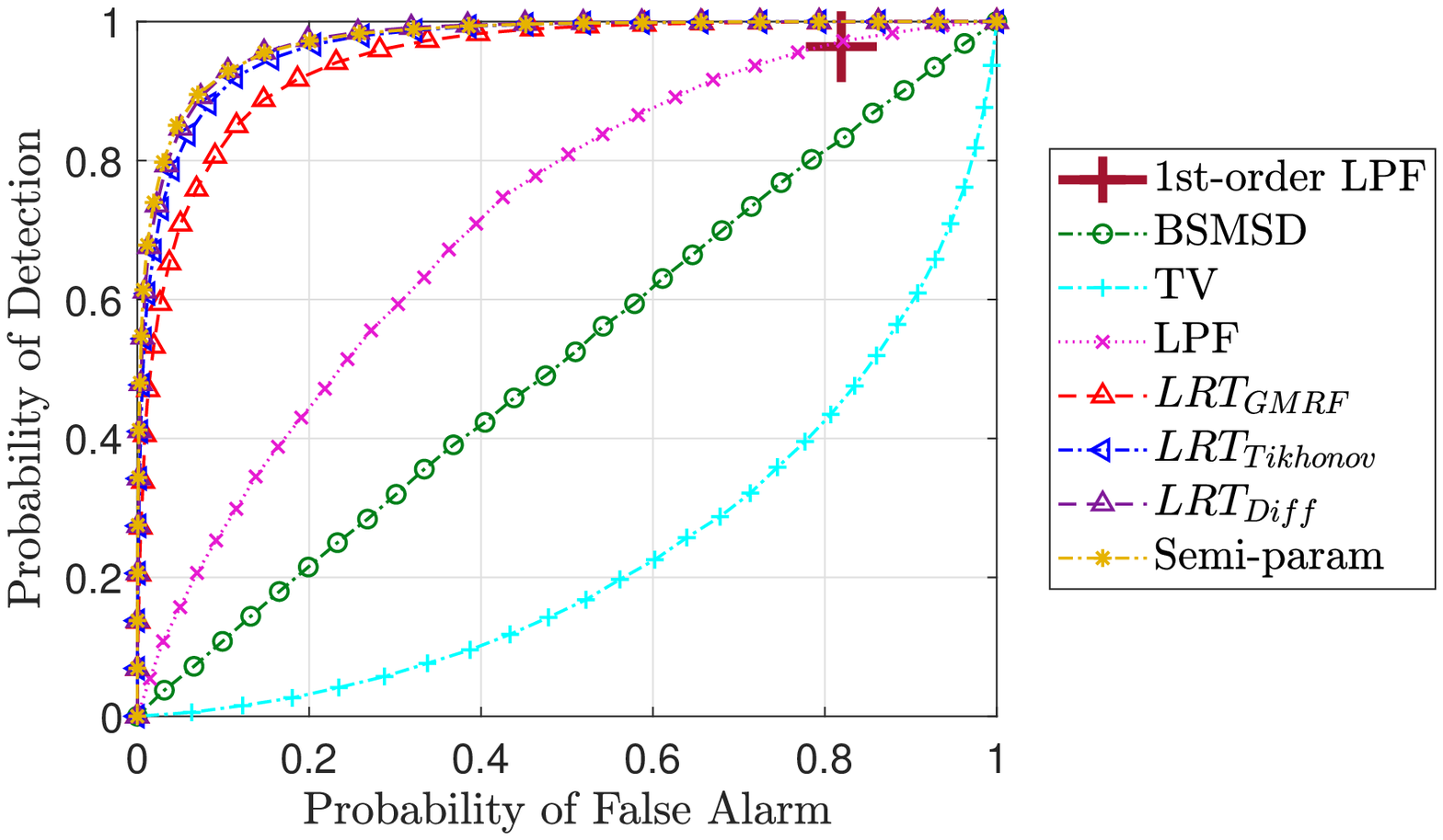}
         \caption{(c)}
         \label{fig:pd_vs_pfa_missmatch}
     \end{subfigure}
        \caption{      The probability of detection of the {different detectors}, where the  probability of false alarm is set to $0.001$, versus (a) the number of samples, (b) $r$.
         In (c) the \ac{roc} curves of the {different detectors} 
 is shown with unequal filter norms under different hypotheses, i.e.  $\text{Tr}(h_1^2(\Lmat))=0.9\text{Tr}(h_0^2(\Lmat))$.}
        \label{fig22}
\end{figure*}

In Fig. \ref{fig:pd_vs_pfa_missmatch}, the \ac{roc} curves of all the detectors 
are  evaluated. The data is generated according to  the  model in \eqref{model}, where $ h_0(\Lmat)=\beta(\Imat+0.2\Lmat)^{-1}$  and $ h_1(\Lmat)=0.9\Imat$. 
 As a result,  $\text{Tr}(h_1^2(\Lmat))=0.89\text{Tr}(h_0^2(\Lmat))$, i.e. the hypotheses differ by both smoothness and scaling of the output graph signals. The results reveal that the \ac{lrt}s, the semi-parametric, the first-order \ac{lpf}, {and the \ac{lpf}} detectors are resilient to scaling, while the BSMSD and the naive \ac{tv} detectors exhibit poor performance in this setting (below the random line of $P_{FA}=P_D$).
The first-order \ac{lpf} has a high probability of false alarm, since it highly depends on the quality of the estimation of the covariance matrix, which is inaccurate due to the small number of samples, $M=30$.

\subsection{Power System Data}\label{subsec;power_sim}
A power system can be represented as an undirected weighted graph, ${\mathcal{G}}({\mathcal{V}},\xi)$, where the set of vertices, $\mathcal{V}$, is the set of buses (generators or loads) and the edge set, $\xi$, is the set of transmission lines  \cite{Giannakis_Wollenberg_2013}.
The Laplacian matrix, $\Lmat$,
is constructed by using the susceptance of the transmission lines \cite{2021Anna,Dabush2021Routtenberg}. 
The voltage data from the power grid
has been shown empirically and theoretically to be smooth/a graph \ac{lpf} signal \cite{Dabush2021Routtenberg,2021Anna,drayer2018detection}.
 In contrast, it is demonstrated empirically that, { although the summation of total real power generations and loads is zero for lossless systems,} the real power signal cannot be considered to be smooth \cite{Dabush2021Routtenberg}. {This can be explained by the fact that each generator/load injects different, uncorrelated power into the system. }
{In the following, we use power data (susceptances, voltage angles, and power measurements under steady-state conditions) obtained from the IEEE 14-bus test case~\cite{iEEEdata}, 
which is a simplified representation of the U.S. Midwest electric power system in 1962, consisting of $N=14$ buses, $5$ generators, and $11$ loads. 
}

We applied the different detectors on the voltage angles vector under $\mathcal{H}_0$, and  on the real power vector under $\mathcal{H}_1$. 
In this scenario, the probability of false alarm refers to the probability of the detectors declaring that the voltage angles are not smooth, while the probability of detection is the probability of declaring the power measurements as not smooth.
The performance is evaluated using a single vector of measurements, $M=1$. 
Gaussian white noise with a covariance matrix $\Rmat=0.2^2\Imat$ was added to the normalized voltage and power vectors. 
Finally, 
for the \ac{lrt} from \eqref{eq:LRT_tikhonov} we set $\alpha=0.02$ and for the \ac{lrt} from \eqref{eq:LRT_diffusion} we set $\tau=0.01$. 
 
Figure \ref{fig:power} presents the \ac{roc} of all the detectors.
The probabilities of detection of the semi-parametric and the \ac{lrt} with the \ac{gmrf} graph filter from \eqref{eq:LRT_GMRF} are the largest for each probability of false alarm.
This can be explained by the similarity of their assumed model, $h_0(\Lmat)=(\Lmat^{0.5})^\dagger$ and $h_1(\Lmat)=\Imat$ (based on the interpretation of the semi-parametric as the \ac{glrt} with unknown variance in Appendix \ref{sub;detectors_partially_known_model}) to the real measurement model, which can be approximated for the voltages and power as $\xvec[m]= \Lmat^\dagger\yvec[m]$ and $\xvec[m]=\yvec[m]$, respectively (based on the linearized power flow model \cite{Giannakis_Wollenberg_2013}).
The first-order \ac{lpf} has a high probability of false alarm, since it highly depends on the quality of the estimation of the covariance matrix, which is inaccurate due to the small number of samples, $M=1$.
The probability of detection of the proposed detectors is significantly higher than those of the BSMSD, the naive \ac{tv},  {and the \ac{lpf}  detectors}, since the BSMSD uses a strict frequency cutoff (here set to $\lambda_7$), which does not fit real voltage signals.
Moreover, the \ac{tv} detector 
heavily relies on data normalization, which is not guaranteed,   { and the \ac{lpf} detector is very sensitive to noise.
}

\subsection{{Wireless Sensor Network Data}}\label{subsec;wsn_sim}
{
From an engineering perspective, data collected from \ac{wsn} is ideally represented by a graph. The \ac{wsn} dataset in the following simulations includes light intensity measurements obtained from distributed sensors within an indoor environment from\cite{intel_research_berkeley_lab},
where only the $40$ most reliable sensors were used.
}
{The associated graph is constructed from the sensor locations, where the edges are determined based on Euclidean distances between sensors $i$ and $j$, similar to Subsection \ref{subsec;synthtic_data_sim}. The edge weights are computed using  \eqref{rbf_edges}, where $d(i, j)$ is the distance between sensors and $\sigma = 10$ is a kernel width. We removed all edges with weights smaller than $0.7$.
}

\begin{figure}[htb]
    \centering
    \subcaptionbox{\label{fig:power}}[\linewidth]
    {\includegraphics[width=6.5cm]{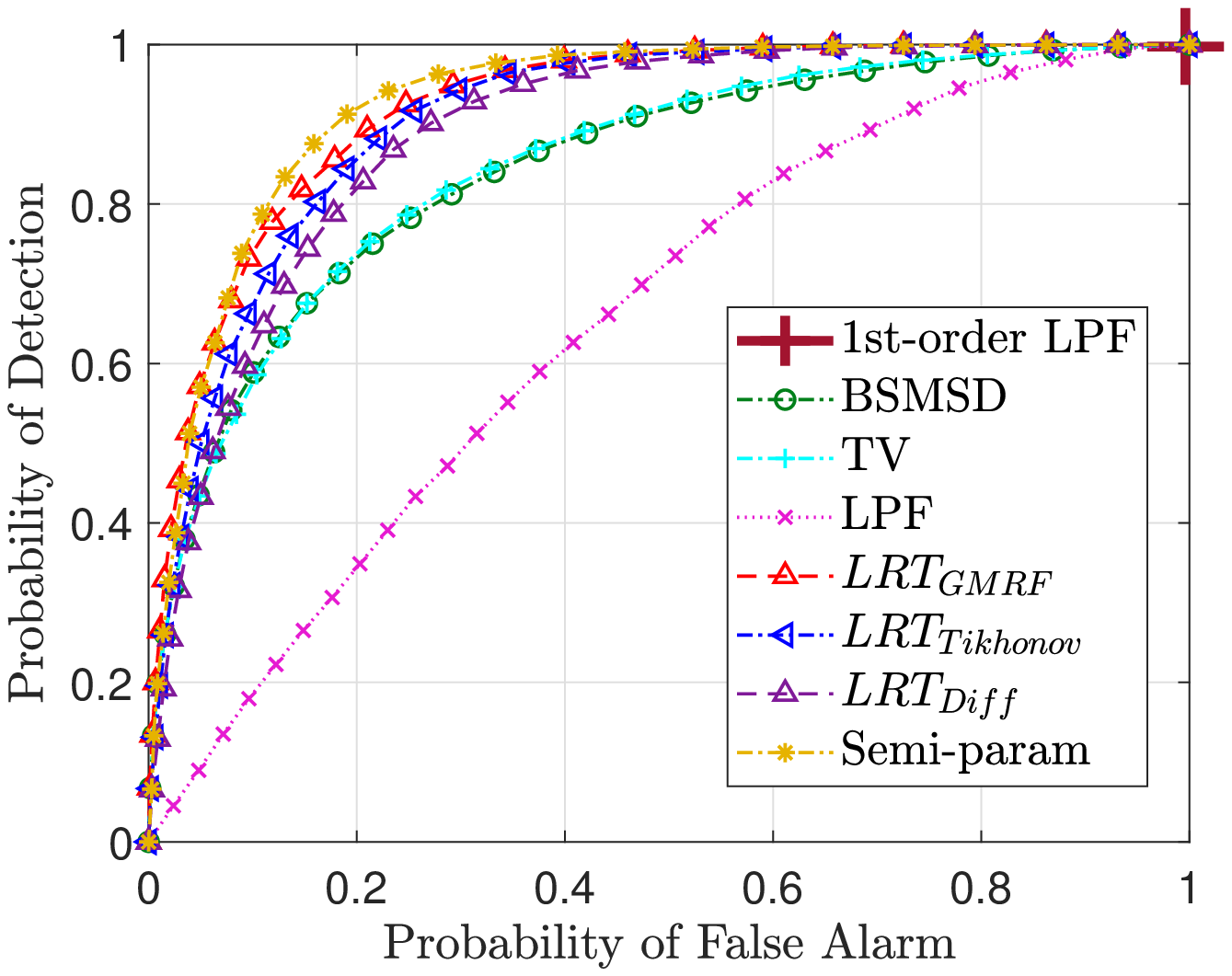}}
    \subcaptionbox{\label{fig:wsn}}[\linewidth]
    {\includegraphics[width=6.5cm]{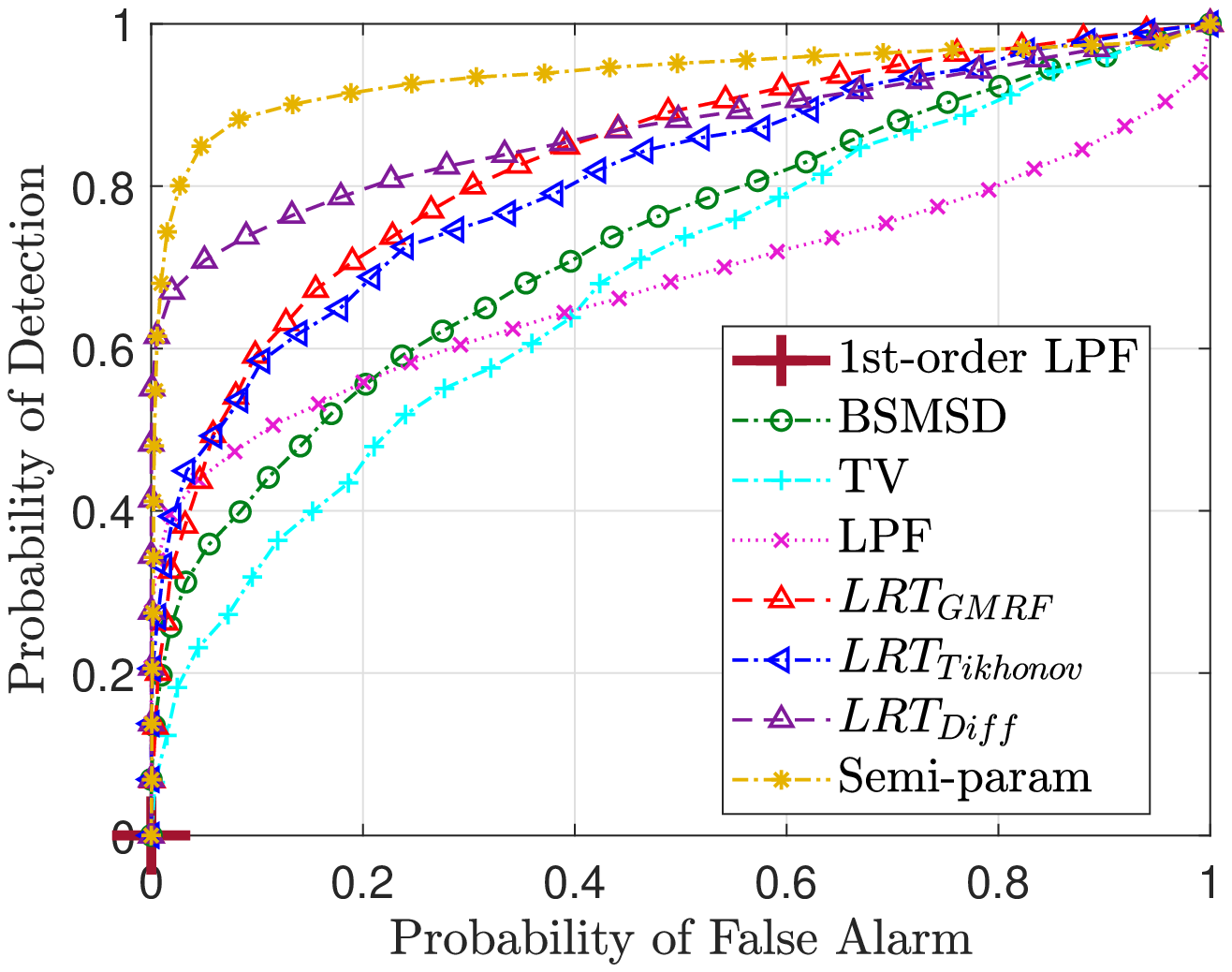}}
    \caption{
   The \ac{roc} curves of the different detectors,
   where under $\mathcal{H}_0$, the measurements are (a)  voltage angles  \cite{iEEEdata} (smooth signal), {(b) light intensity \cite{intel_research_berkeley_lab},} and under $\mathcal{H}_1$, the measurements are (a)  real power data \cite{iEEEdata} (non-smooth signal), {(b) a random permutation of the measurements.}}
     \label{fig:PSSE}
\end{figure}

{
Light intensity typically decays approximately as $1/R^2$, where $R$ is the distance to the light source, resulting in a smooth distribution of light intensity. However, due to the presence of walls and obstructing objects, the smoothness of the light intensity may be destroyed. Thus, this property is system-dependent and needs to be validated.
Figure \ref{fig:wsn}, 
shows the \ac{roc} curves of all the detectors where the measurement vectors are the light intensity under $\mathcal{H}_0$  and a random permutation of these measurements under $\mathcal{H}_1$, which is not smooth \ac{wrt} the original graph. 
Under both hypotheses, the number of samples is $M=40$, and zero-mean Gaussian white noise with $\sigma=1$ was added. 
The detectors' parameters were tuned based on humidity measurements also obtained from these sensors \cite{intel_research_berkeley_lab}, where we obtained $\alpha=20$, $\tau=10$, and $k=7$ for the Tikhonov Regularization \ac{lrt} from \eqref{eq:LRT_tikhonov}, for the Diffusion Kernel \ac{lrt} from \eqref{eq:LRT_diffusion}, and for the \ac{lpf} detector from \eqref{eta_k_def_est_lpf}, respectively.}
{{
It can be seen that the detectors identify the light intensity as a smooth graph signal, where
the semi-parametric method exhibits superior performance in distinguishing light measurements from randomly permuted measurements.  The first-order \ac{lpf}, which is not aware of the graph structure, cannot distinguish between the original data and its permutation, resulting in zero probabilities of detection and false alarm.
The differences between the \ac{lrt}s with the different filters stem from how well they fit the data. 
}}

\section{Conclusion}\label{Conclusions}
In this paper we investigate the validation of the smoothness assumption \ac{wrt} a given graph topology.
We formulate the detection of smooth graph signals as a hypothesis-testing problem 
with a graph-filtered signal model. 
We show that covariance-based detectors are capable of capturing the signal smoothness under this model. In addition, we propose a new definition of smooth graph filters and derive its analytical relation with the graph \ac{lpf}s \cite{Rama2020Anna}. 
 We propose 
the \ac{lrt} and the semi-parametric smoothness detectors, and analyze their properties under different scenarios and in the graph spectral domain.
It is shown that the semi-parametric detector can be interpreted as the \ac{glrt} for specific known graph filters (inverse Laplacian and all-pass filters) with a Gaussian input graph signal with an unknown variance under the two hypotheses.
We show how to set the threshold of the proposed \ac{lrt} and semi-parametric detectors by providing analytical/approximated terms for the associated probability of false alarm.
The proposed methods require less data than the first-order graph \ac{lpf} method, and do not assume an ideal graph \ac{lpf} as 
the BSMSD. 
Simulation results 
demonstrate that the proposed methods outperform existing methods on the tested scenarios in terms of
detection performance. {In particular, they demonstrate superior performance with a larger scale of smoothness, compared to other methods,  
require a smaller dataset, especially in comparison with the first-order \ac{lpf}, and are robust to scaling, unlike the BSMSD and the naive \ac{tv} detector.} Moreover,
 the semi-parametric detector {exhibits the lowest computational complexity,} is consistent with the \ac{lrt}s for synthetic data, and exhibits the highest level of accuracy for power data. Its ability to provide good performance without relying on specific graph filters or the use of tuning parameters makes it an attractive detector.

 \appendices
	\renewcommand{\thesectiondis}[2]{\Alph{section}:}
\section{Proof of Theorem \ref{non_increasing_theorem}}\label{appendix_smoothness_of_box_filters}
 {
According to Chebyshev's sum inequality
(\hspace{-0.01cm}\cite{hardy1952inequalities}, pp. 43-44),  if $\{a_n\}_{n=1}^N$ are non-increasing and $\{b_n\}_{n=1}^N$ are non-decreasing sequences, then 
\begin{equation}\label{chebyshev}
\sum_{n=1}^Na_nb_n < \frac{1}{N}\left(\sum_{n=1}^Na_n\right)\left(\sum_{n=1}^Nb_n\right),    
\end{equation}
unless all the $\{a_n\}_{n=1}^N$ or all the $\{b_n\}_{n=1}^N$ are equal.}
 {
 Thus, for any non-increasing graph filter, $h(\cdot)$, with at least two distinct coefficients $i$ and $j$ such that $h(\lambda_i) \neq h(\lambda_j)$, the sequences 
 $a_n=\frac{\lambda_n}{\sum_{n=1}^N\lambda_n}$ and $b_n=\frac{{h^2({\lambda}_n)}}{\sum_{n=1}^N{h^2({\lambda}_n)}}$ are 
non-decreasing  and
 non-increasing sequences, respectively, for $n=1,\ldots,N$, and $\sum_{n=1}^Na_n=\sum_{n=1}^Nb_n=1$.
 By substituting these sequences in \eqref{chebyshev} and using the fact that $\frac{r}{N}$ from \eqref{eq_SGF_gft} 
 can be written as $\sum_{n=1}^Na_nb_n$ for these sequences,   we obtain that 
$\frac{r}{N}<\frac{1}{N}$.}

	\section{Proof of Claim \ref{claim_equiv_to_lpf} }\label{proof_claim_equiv_to_lpf} 
 First, we note that the condition in \eqref{eq_SGF_gft} can be rewritten as 
  \beqna\label{eq_SGF_gft2}
     \sum\nolimits_{n=1}^N\bar{\lambda}_n h^2({\lambda}_n)<0, 
     \eeqna  
     where in this appendix we use the notation $\bar{\lambda}_n=\lambda_n-\lambda_{\rm{avg}}$. 
To ensure that any graph filter $h(\Lmat)$ that satisfies the conditions in Claim \ref{claim_equiv_to_lpf} satisfies \eqref{eq_SGF_gft2}, we need to show that the maximum of the l.h.s. of \eqref{eq_SGF_gft2} w.r.t.  $h(\Lmat)$ that
satisfies these conditions is negative. 
Thus, we define the following maximization:
  \beqna\label{eq_goal_of_proof}
      \max_{h(\pmb{\lambda}) \in {\mathbb{R}}^N}
      \sum_{n=1}^N{\bar{\lambda}_n h^2({\lambda}_n)}
     ~~~ \text{s.t.}\quad 
      \eta_K^2<\frac{\sum_{n=1}^K\bar{\lambda}_n}{\sum_{n=1}^J\bar{\lambda}_n},
      \eeqna  
      where $K\leq J$ and
      the constraint in \eqref{eq_goal_of_proof}
      is the condition in \eqref{eq_cond_eta_1}.
By substituting \eqref{eta_k_def} in \eqref{eq_goal_of_proof} and splitting the sum in the objective of \eqref{eq_goal_of_proof},
the problem in \eqref{eq_goal_of_proof} is equivalent to 
  \beqna\label{eq_goal_of_proof2}
\max_{h(\pmb{\lambda}) \in {\mathbb{R}}^N}
\sum_{n=1}^J\bar{\lambda}_nh^2(\lambda_n)
+\sum_{n=J+1}^N\bar{\lambda}_n h^2(\lambda_n)\hspace{1.5cm}\nonumber\\
      \text{s.t.}  \quad
       \frac{\max\{{h}^2({\lambda}_{K+1}),\ldots,{h}^2({\lambda}_{N})\}}{\min\{{h}^2({\lambda}_{1}),\ldots,{h}^2({\lambda}_{K})\}}<\frac{\sum_{n=1}^K\bar{\lambda}_n}{\sum_{n=1}^J\bar{\lambda}_n}.
      \eeqna  
Since for $n\leq J$ we know that $\bar{\lambda}_n\leq 0$, we can bound the left sum in the objective function in \eqref{eq_goal_of_proof2}
as follows:
\beqna\label{eq:negative_bound}
\sum\nolimits_{n=1}^J\bar{\lambda}_n h^2(\lambda_n)\leq \sum\nolimits_{n=1}^K \bar{\lambda}_n h^2(\lambda_n)\nonumber\\\leq\min\{h^2(\lambda_1),\ldots,h^2(\lambda_K)\}\sum\nolimits_{n=1}^K\bar{\lambda}_n.
\eeqna
Similarly, for $n> J$ we know that $\bar{\lambda}_n> 0$,  and thus, 
\beqna\label{eq:positive_bound}
\sum\nolimits_{n=J+1}^N\bar{\lambda}_n h^2(\lambda_n)\hspace{3cm}\nonumber\\\leq\max\{h^2(\lambda_{J+1}),\ldots,h^2(\lambda_N)\}\sum\nolimits_{n=J+1}^N\bar{\lambda}_n \nonumber\\
 \leq\max\{h^2(\lambda_{K+1}),\ldots,h^2(\lambda_N)\}\sum\nolimits_{n=J+1}^N\bar{\lambda}_n,
\eeqna 
where the last inequality is since we added elements to the maximization. 
By substituting  $\sum_{n=J+1}^N\bar{\lambda}_n=-\sum_{n=1}^J\bar{\lambda}_n$, which arises from the properties of the average
 \eqref{eq:positive_bound}, we obtain
\beqna\label{eq:positive_bound2}
\sum\nolimits_{n=J+1}^N\bar{\lambda}_n h^2(\lambda_n)\hspace{3.5cm}\nonumber\\
 \leq\max\{h^2(\lambda_{K+1}),\ldots,h^2(\lambda_N)\}\cdot (-\sum\nolimits_{n=1}^J\bar{\lambda}_n).
\eeqna 
By substituting \eqref{eq:negative_bound} and \eqref{eq:positive_bound2} in \eqref{eq_goal_of_proof2}, we obtain 
  \beqna\label{eq_goal_of_proof3}    \max_{h(\pmb{\lambda}) \in {\mathbb{R}}^N}
  \Big\{\min\{{h}^2({\lambda}_{1}),\ldots,{h}^2({\lambda}_{K})\}\sum\nolimits_{n=1}^K\bar{\lambda}_n\hspace{0.1cm}\nonumber\\
	    +\max\{h^2(\lambda_{K+1}),\ldots,h^2(\lambda_N)\}(-\sum\nolimits_{n=1}^J\bar{\lambda}_n)\Big\}\nonumber\\
      \text{s.t.} 
       \frac{\max\{{h}^2({\lambda}_{K+1}),\ldots,{h}^2({\lambda}_{N})\}}{\min\{{h}^2({\lambda}_{1}),\ldots,{h}^2({\lambda}_{K})\}}<\frac{\sum_{n=1}^K\bar{\lambda}_n}{\sum_{n=1}^J\bar{\lambda}_n}.
      \eeqna  
The result of the optimization in \eqref{eq_goal_of_proof3} is always larger than or equal to the solution of \eqref{eq_goal_of_proof2}. Thus, by showing that the optimal solution of \eqref{eq_goal_of_proof3} is negative, we immediately obtain that the optimal solution of \eqref{eq_goal_of_proof2} is negative. 
By dividing the objective in  \eqref{eq_goal_of_proof3} by 
$\min\{{h}^2({\lambda}_{1}),\ldots,{h}^2({\lambda}_{K})\}$, one obtains
\beqna\label{eq_goal_of_proof4}
\max_{h(\pmb{\lambda}) \in {\mathbb{R}}^N}
\Big\{\sum_{n=1}^K\bar{\lambda}_n\hspace{4.4cm}\nonumber\\+ \frac{\max\{{h}^2({\lambda}_{K+1}),\ldots,{h}^2({\lambda}_{N})\}}{\min\{{h}^2({\lambda}_{1}),\ldots,{h}^2({\lambda}_{K})\}}\Big(-\sum_{n=1}^J\bar{\lambda}_n\Big)\Big\}\nonumber\\
      \text{s.t.} 
       \frac{\max\{{h}^2({\lambda}_{K+1}),\ldots,{h}^2({\lambda}_{N})\}}{\min\{{h}^2({\lambda}_{1}),\ldots,{h}^2({\lambda}_{K})\}}<\frac{\sum_{n=1}^K\bar{\lambda}_n}{\sum_{n=1}^J\bar{\lambda}_n}.
      \eeqna  
Finally, by using the constraint from \eqref{eq_goal_of_proof4}, the objective in  \eqref{eq_goal_of_proof4} can be bounded from above as follows:
\beqna\label{eq_seperated_sum4}
\Big\{\sum_{n=1}^K\bar{\lambda}_n+ \frac{\max\{{h}^2({\lambda}_{k+1}),\ldots,{h}^2({\lambda}_{N})\}}{\min\{{h}^2({\lambda}_{1}),\ldots,{h}^2({\lambda}_{k})\}}\Big(-\sum_{n=1}^J\bar{\lambda}_n\Big)\Big\}\nonumber\\
<\sum\nolimits_{n=1}^K\bar{\lambda}_n
	    -\sum\nolimits_{n=1}^J\bar{\lambda}_n\frac{\sum\nolimits_{n=1}^K\bar{\lambda}_n}{\sum\nolimits_{n=1}^J\bar{\lambda}_n})=0.
	\eeqna
The result in \eqref{eq_seperated_sum4} shows that the objective in \eqref{eq_goal_of_proof} is negative; Hence, a graph filter that satisfies \eqref{eq_cond_eta_1} with $K\leq J$ is smooth.

\section{Proof of Claim \ref{prop_mu}}\label{proof_prop_mu}
First, we show that ${\rm{E}}[\bar{\xvec}^T[m]{\Lmat}(\bar{\xvec}[m])]={\rm{E}}[{\xvec}^T[m]{\Lmat}({\xvec}[m])]$.
For the model in \eqref{model_mu}, ${\rm{E}}[\bar{\xvec}[m]]=\pmb{\mu}$ and we can write
\begin{align}\label{eq_expected_smoothness_mu}
{\rm{E}}[\bar{\xvec}^T[m]{\Lmat}(\bar{\xvec}[m])]\hspace{4.5cm}
\nonumber\\={\rm{E}}[(\bar{\xvec}[m]-\pmb{\mu})^T\Lmat(\bar{\xvec}[m]-\pmb{\mu})]+{\rm{E}}[\pmb{\mu}^T\Lmat\pmb{\mu}].
\end{align}
Utilizing the trace properties,  i.e. $\text{Tr}(\Amat\Bmat)=\text{Tr}(\Bmat\Amat)$,  and  the fact that $\bar{\xvec}[m]\sim\mathcal{N}(\pmb{\mu},\sigma^2\bar{h}^2(\Lmat))$, we get
\begin{align}\label{eq_expected_smoothness_x_minus_mu2}
{\rm{E}}[(\bar{\xvec}[m]-\pmb{\mu})^T\Lmat(\bar{\xvec}[m]-\pmb{\mu})]\hspace{3.5cm}\nonumber\\
=\text{Tr}(\Lmat {\rm{E}}[(\bar{\xvec}[m]-\pmb{\mu})(\bar{\xvec}[m]-\pmb{\mu})^T])=\sigma^2\text{Tr}(\Lmat \bar{h}^2(\Lmat)).
\end{align}
By substituting \eqref{eq_expected_smoothness_x_minus_mu2} in \eqref{eq_expected_smoothness_mu} one obtains 
\begin{align}\label{eq_expected_smoothness_model_mu}
{\rm{E}}[\bar{\xvec}^T[m]{\Lmat}(\bar{\xvec}[m])]=\sigma^2{Tr}(\Lmat \bar{h}^2(\Lmat))+{\rm{E}}[\pmb{\mu}^T\Lmat\pmb{\mu}].
\end{align}
For the model in \eqref{model}-\eqref{filter_mu}, ${\xvec}[m]\sim\mathcal{N}(\zerovec,\sigma^2{h}^2(\Lmat))$; therefore, by using the properties of the trace operator,
we obtain
\beqna\label{eq_expected_smoothness_x_bar}
{\rm{E}}[{\xvec}^T[m]\Lmat{\xvec}[m]]=\text{Tr}( {\rm{E}}[{\xvec}^T[m]\Lmat{\xvec}[m])\hspace{2.3cm}\nonumber\\
=\text{Tr}(\Lmat {\rm{E}}[{\xvec}[m]\bar{\xvec}^T[m]])=\sigma^2\text{Tr}(\Lmat {h}^2(\Lmat)).
\eeqna
Substituting the definition of ${h}^2(\Lmat)$ from \eqref{filter_mu} in \eqref{eq_expected_smoothness_x_bar}, we get
\beqna\label{eq_expected_smoothness_x_bar2}
{\rm{E}}[{\xvec}^T[m]\Lmat{\xvec}[m]]=\sigma^2\text{Tr}(\Lmat (\bar{h}^2(\Lmat)+\frac{1}{\sigma^2}\pmb{\mu}\pmb{\mu}^T))\hspace{0.1cm}\nonumber\\
=\sigma^2\text{Tr}(\Lmat \bar{h}^2(\Lmat))+{\rm{E}}[\pmb{\mu}^T\Lmat\pmb{\mu}].
\eeqna
The terms in \eqref{eq_expected_smoothness_model_mu} and \eqref{eq_expected_smoothness_x_bar2} are equal.

The proof of the second part of the claim is composed of two steps: 1) Proof that if $h(\Lmat)$ is symmetric, it has the same eigenvectors as $h^2(\Lmat)$;
 2) Proof that if $\pmb{\mu}$ is proportional to an eigenvectors of the Laplacian, then $\Vmat^{-1} {h}(\Lmat)\Vmat$ is diagonal. 

{\bf{Step 1:}} Given a symmetric matrix $h(\Lmat)$, then, according to  \eqref{filter_mu}, $h^2(\Lmat)$ is also symmetric. Since $h(\Lmat)$ and $h^2(\Lmat)$ are symmetric and commute, they are simultaneously diagonalizable (see Theorem 1.3.12 in \cite{Horn2012}), and share the same eigenvectors.

{\bf{Step 2:}} For ${h}(\Lmat)$ from \eqref{filter_mu} to be a graph filter, $\Vmat^{-1} {h}(\Lmat)\Vmat$ should be a diagonal matrix.
Since $\bar{h}(\Lmat)$ is a valid graph filter, then,  according to the definition in \eqref{graph_filter}, we obtain that 
$\Vmat^{-1}\bar{h}^2(\Lmat)\Vmat$  is a diagonal matrix. 
One can observe that  $(\pmb{\mu}\pmb{\mu}^T)\uvec=\pmb{\mu}(\pmb{\mu}^T\uvec)=(\pmb{\mu}^T\uvec)\pmb{\mu}$ for any vector $\uvec$,
i.e. the vector obtained  by multiplying  $\pmb{\mu}\pmb{\mu}^T$ is proportional to $\pmb{\mu}$. Thus, if $\uvec\propto\pmb{\mu}$, then $\uvec$ is an eigenvector of $\pmb{\mu}\pmb{\mu}^T$, because $\pmb{\mu}\pmb{\mu}^T\pmb{\mu}=||\pmb{\mu}||^2\pmb{\mu}$. If $\pmb{\mu}$ is an eigenvector of $\Vmat$, then, for the other eigenvectors of $\Vmat$ that are orthogonal to $\pmb{\mu}$, $(\pmb{\mu}\pmb{\mu}^T)\uvec=\zerovec$; thus, they are eigenvectors of $\pmb{\mu}\pmb{\mu}^T$ that correspond to eigenvalues equal to zero. Thus, if $\pmb{\mu}$ is equal to one of the columns of $\Vmat$, $\Vmat^{-1}\bar{h}^2(\Lmat)\Vmat$  is a diagonal matrix.

\section{Proof of Proposition \ref{prop}}\label{Performance_Analysis}
By using the \ac{gft} definition from \eqref{GFT}, we obtain
\begin{equation}\label{eq_gft_general_test}
\sum\nolimits_{m=1}^M \xvec^T[m] h_{a}(\Lmat)\xvec[m]=\sum\nolimits_{m=1}^M\sum\nolimits_{n=1}^{N}h_{a}(\lambda_n)\tilde{x}_n^2[m].
\end{equation}
Under the assumption in Proposition \ref{prop}  that $\{\xvec[m]\}_{m=1}^M$ is an \ac{iid} sequence with  $\xvec[m]\sim \mathcal{N}(\zerovec,h_{b}^2(\Lmat))$,  $\tilde{x}_n[m]$
is also Gaussian. 
From the linearity of the \ac{gft} in \eqref{GFT}, we obtain  that $
\text{E}[\tilde{\xvec}[m]]=
  \Vmat^{-1} \text{E}[{\xvec}[m]]=\zerovec$,
and the covariance matrix is
\beqna\label{eq_transformed_cov}
  \text{E}[\tilde{\xvec}[m]\tilde{\xvec}^T[m]]=\Vmat^{-1}\text{E}[{\xvec[m]}{\xvec}^T[m]]\Vmat=\text{diag}(h_{b}^2(\pmb{\lambda})).
\eeqna
Hence, 
$\tilde{\xvec}[m]|~\mathcal{H}_0\sim\mathcal{N}\left(\zerovec,\text{diag}(h_{b}^2(\pmb{\lambda}))\right)$, $m=1,\dots,M$.

The entries of $\tilde{\xvec}[m]$ correspond to the indices where $h_{b}(\pmb{\lambda})=0$ are deterministic zero parameters.
Therefore, 
\beqna\label{eq_pfa_general_appendix}
   \sum_{m=1}^M\sum_{n=1}^N h_{a}(\lambda_n)\tilde{x}_n^2[m]
= \sum_{m=1}^M\sum_{n\in {\text{supp}}(h_b)} h_{a}(\lambda_n)\tilde{x}_n^2[m]\nonumber\\= \sum_{n\in {\text{supp}}(h_b)}h_a(\lambda_n)h_b^2(\lambda_n)
 \sum_{m=1}^M\frac{\tilde{x}_n^2[m]}{h_b^2(\lambda_n)}.
\eeqna
From the distribution of $\tilde{\xvec}[m]$ under hypothesis $\mathcal{H}_0$, 
we have 
$ \sum_{m=1}^M\tilde{x}_n^2[m]/h_b^2(\lambda_n)$, $ n \in {\text{supp}}(h_b)$, are independent centralized $\chi$-square variables with $M$ degrees of freedom.
Thus, by substituting the weights from \eqref{eq_weights_general} in \eqref{eq_pfa_general_appendix}, 
\beqna\label{eq_Q_def}
\Pr\left(\sum\nolimits_{n\in {\text{supp}}(h_b)}
w(\lambda_n)
 \sum\nolimits_{m=1}^M\frac{\tilde{x}_n^2[m]}{h_b^2(\lambda_n)}\leq\gamma\right)
 \nonumber\\
 =
 Q({M,\{w(\lambda_n)\}_{n \in {\text{supp}}(h_b)}},\gamma).
\eeqna 
From \eqref{eq_Q_def} we obtain \eqref{eq_pfa_general}, where $w(\lambda_n)$ is defined in \eqref{eq_weights_general}.


\section{Derivation of \eqref{opt_A} and \eqref{opt_A2}}\label{appendix_filter_coeff}
In this appendix we derive the constrained  \ac{ml} estimator of $\sigma^2h^2(\lambda_n)$ based on the log-likelihood in \eqref{log_semi_gft} under the constraint from \eqref{constraint1}.
We denote by ${\mathcal{S}}_n=\{j|\lambda_j=\lambda_n\}$ the set of the indices of all eigenvalues that are equal to $\lambda_n$.
To simplify the derivation we substitute  $a_n\define\sigma^2h^2(\lambda_n)$ and 
\beqna\label{coeff_def_gen}
c_n\define\frac{1}{M}\sum\nolimits_{m=1}^M\tilde{\xvec}_n^2[m], ~~n\in\mathcal{D},
\eeqna
in \eqref{log_semi_gft} and perform the maximization according to $a_n$:
\beqna
\label{eq:Smooth_wls_ac}
\hat{a}_1,\ldots,\hat{a}_N=\arg\min_{a_1,\ldots,a_N\in\mathbb{R}}\sum_{n=1}^N \log(a_n^{-1})-c_na_n^{-1}
\nonumber\\
{\text{s.t.  }} a_m = a_k, \forall m\not=k,~m,k\in\mathcal{S}_K.
\eeqna
The Lagrangian of this minimization problem is
\beqna
\sum_{n=1}^N (\log(a_n^{-1})-c_na_n^{-1})+\sum_{m\not=K,m \in{\mathcal{S}}_k} \rho_m(a_m-a_K),
\eeqna
where $\rho_m,~\forall m\in\mathcal{S}_K$ are the Lagrange multipliers.
By using the \ac{kkt} conditions, we get
\beqna
0&=&-a_i^{-1}+c_ia_i^{-2}+\rho_i, ~\forall i\not=K,i\in\mathcal{S}_K,
\\
 0&=&-a_i^{-1}+c_ia_i^{-2} -\sum_{m\not=K,m \in{\mathcal{S}}_K}\rho_m, ~i=K,
\\
 0&=&-a_i^{-1}+c_ia_i^{-2},~\forall i\not\in\mathcal{S}_K,
\\
 a_i&=&a_K,~\forall i\in\mathcal{S}_K.
\eeqna
After some mathematical manipulations, we get
\beqna
\label{estimator}
\hat{a}_n=\hat{\sigma}^2\hat{h}^2(\lambda_n)\left\{\begin{array}{l}0,\quad n\not\in\mathcal{D}\\
c_n,\quad n\not\in\mathcal{S}_K\\
\frac{1}{|\mathcal{S}_K|}\sum_{m\in\mathcal{S}_K}c_m,~~n\in\mathcal{S}_K
\end{array}\right..
\eeqna
Substituting \eqref{coeff_def_gen} in \eqref{estimator} yields \eqref{opt_A} and \eqref{opt_A2}.
\section{Detectors based on a partially known model under both hypotheses}\label{sub;detectors_partially_known_model}
Assume the hypothesis testing in \eqref{eq:hypothesis2},
where $h_{0}(\Lmat)$ and $h_{1}(\Lmat)$ are known,  while $\sigma^2$ is unknown.
The problem in \eqref{eq:hypothesis2} is a composite hypothesis testing problem, since the 
measurements depend on the unknown variance $\sigma^2$. Hence, the \ac{glrt} is used here, where 
$\sigma^2$ in the \ac{lrt} is replaced by its
  \ac{ml} estimator (Chapter 6.4 in \cite{Kay_detection}).
Under $\mathcal{H}_i$, the log-likelihood is given in \eqref{log_trans}, and the \ac{ml} estimator of $\sigma^2$ is 
\beqna
\label{ml_sigma1}
\hat{\sigma^2}_{\mathcal{H}_i}=\frac{1}{MN}
\sum_{m=1}^M\xvec^T[m](h_i^2(\Lmat))^{\dagger}\xvec[m],~~i=\{0,1\}.
\eeqna
By substituting \eqref{ml_sigma1}  in the log-likelihood and ignoring constant terms, the \ac{glrt} after applying $\exp(\cdot)$ on both sides is 
\begin{equation}\label{eq_glrt_partially_known}
    \text{GLRT}(\xvec)=\frac{\sum_{m=1}^M\xvec[m]^T(h_{0}^2(\Lmat))^{\dagger}\xvec[m]}{\sum_{m=1}^M\xvec^T[m](h_{1}^2(\Lmat))^{\dagger}\xvec[m]} \mathop{\gtrless}^{H_1}_{H_0}\,\gamma.
\end{equation}
For $h_{0}(\Lmat)=\Lmat^{\dagger}$ and $h_{1}(\Lmat)=\Imat$ we receive the semi-parametric detector from \eqref{eq_semi_detector2}.

The probability of false alarm of the \ac{glrt} derived for the hypotheses testing in \eqref{eq:hypothesis2}  from \eqref{eq_glrt_partially_known} is 
    \beqna
    \label{Pr_error}
  \Pr\left(  \text{GLRT}(\xvec)>\gamma\right)
   \hspace{5.4cm}\nonumber\\=   
     \Pr\left(\frac{\sum_{m=1}^M 1/\sigma^2\xvec^T[m] (h_{0}^2(\Lmat))^{\dagger}\xvec[m]}{\sum_{m=1}^M 1/\sigma^2\xvec^T[m] (h_{1}^2(\Lmat))^{\dagger}\xvec[m]}>\gamma\right)
   \hspace{1.1cm}\nonumber\\=
   \Pr\left(\sum_{m=1}^M \xvec^T[m] 1/\sigma^2((h_{0}^2(\Lmat))^{\dagger}-\gamma (h_{1}^2(\Lmat))^{\dagger})\xvec[m]>0\right)\hspace{-0.6cm}\nonumber\\
   =1-Q(M,\{h_{0}^2(\Lmat)((h_{0}^2(\Lmat))^{\dagger}-\gamma (h_{1}^2(\Lmat))^{\dagger})\}_{n=1}^N,0),\hspace{0.2cm}
    \eeqna
    where the last equality is obtained by substituting 
    $h_a(\Lmat)=1/\sigma^2((h_{0}^2(\Lmat))^{\dagger}-\gamma (h_{1}^2(\Lmat))^{\dagger})$ and $h_b^2(\Lmat)=\sigma^2h_{0}^2(\Lmat)$  in the weights from \eqref{eq_weights_general} according to Proposition \ref{prop}. Since the false alarm probability in \eqref{Pr_error} is independent of $\sigma^2$, and thus, it can be calculated despite the fact that $\sigma^2$ is unknown. 
\bibliographystyle{IEEEtran}
\bibliography{IEEEabrv,main}

\end{document}